\documentclass[reprint,showpacs,preprintnumbers,amssymb,superscriptaddress,aps,prd]{revtex4-1}

\usepackage{graphicx, epsfig, amssymb} 
\usepackage{amsmath, amsfonts}
\usepackage{bm}

\usepackage[linktocpage]{hyperref}
\usepackage[usenames]{color}
\usepackage{subfigure}
\usepackage[utf8]{inputenc}
\usepackage{tensor}
\usepackage{soul}

\def\b{\begin{equation}}
\def\e{\end{equation}}

\def\be{\begin{eqnarray}}
\def\ee{\end{eqnarray}}

\newcommand{\Veff}{V_{\text{eff}}}

\newcommand{\ARNtra}{A_{RN}^{\text{tra}}\,}
\newcommand{\ARNinc}{A_{RN}^{\text{inc}}\,}
\newcommand{\ARNref}{A_{RN}^{\text{ref}}\,}
\newcommand{\Aetra}{A_e^{\text{tra}}}
\newcommand{\Aeinc}{A_e^{\text{inc}}}
\newcommand{\Aeref}{A_e^{\text{ref}}}
\newcommand{\sighf}{\sigma_{\text{hf}}}
\newcommand{\siglf}{\sigma_{\text{lf}}}
\newcommand{\FF}{F}
\newcommand{\sinc}{\text{sinc}}
\newcommand{\mass}{m}
\newcommand{\Mm}{M\mass}

\begin{document}\title{\large Absorption of a massive scalar field by a charged black hole}

\author{Carolina L. Benone}\email{lben.carol@gmail.com}
\affiliation{Faculdade de F\'{\i}sica, Universidade Federal do Par\'a, 66075-110, Bel\'em, Par\'a, Brazil}

\author{Ednilton S. de Oliveira}\email{esdeoliveira@gmail.com}
\affiliation{Faculdade de F\'{\i}sica, Universidade Federal do Par\'a, 66075-110, Bel\'em, Par\'a, Brazil}

\author{Sam R. Dolan}\email{s.dolan@sheffield.ac.uk}
\affiliation{Consortium for Fundamental Physics,
School of Mathematics and Statistics,
University of Sheffield, Hicks Building, Hounsfield Road, Sheffield S3 7RH, United Kingdom}

\author{Lu\'{\i}s C. B. Crispino}\email{crispino@ufpa.br}
\affiliation{Faculdade de F\'{\i}sica, Universidade Federal do Par\'a, 66075-110, Bel\'em, Par\'a, Brazil}

\begin{abstract}
We calculate the absorption cross section of a massive neutral scalar field impinging upon a Reissner-Nordstr\"om black hole. First, we derive key approximations in the high- and low-frequency regimes. Next, we develop a numerical method to compute the cross section at intermediate frequencies, and present a selection of results. Finally, we draw together our complementary approaches to give a quantitative full-spectrum description of absorption.
\end{abstract}

\date{\today}

\maketitle

\section{Introduction\label{sec:intro}}

It is almost a century since the theory of General Relativity (GR) supplanted Newton's theory as the leading explanation for gravitational phenomena. GR was quickly recognized as bearing the hallmarks of a successful theory, because not only was it consistent with the existing canon of data, but it also resolved known anomalies (such as the anomalous precession of Mercury noted by Le Verrier in 1859), and made new predictions which soon passed experimental tests, such as the deflection of starlight measured by Eddington's eclipse expedition.

Just as importantly, GR provided a mathematically-consistent extension of the concepts of special relativity, which allowed gravitation to be reinterpreted as a consequence of the geometry of spacetime. This reinterpretation instigated a revolution in our understanding of the Universe. GR is the framework underpinning structure formation in an expanding, and accelerating, Cosmos. GR also provides radical `strong-field' predictions which test the theory to its limits, namely, the black holes (BHs). 
Stationary BHs are simple solutions of Einstein's equations, that (in electrovacuum) depend only on three parameters: mass, charge, and angular momentum \cite{Bardeen:1973}. On the other hand, dynamical BHs in astrophysical environments are undoubtedly crucibles for tests of physics, including (in principle) the unification of GR and quantum theory.

BHs may be classified by mass, into three categories \cite{Celotti:1999tg,Narayan:2013gca}: primordial BHs, formed in the early Universe; stellar-mass BHs, formed after the death of stars; and super-massive BHs, formed in the center of galaxies. Since the development of X-ray astronomy in the 1970s, there has been an accumulation of very strong indirect evidence for the existence of stellar-mass and supermassive BHs. The existence of primordial BHs remains open to speculation \cite{Green:2014faa}; it is thought likely that, by the present epoch, all primordial BHs with initial masses $\lesssim 10^{12}$kg have evaporated via Hawking emission.

The spacetime of static charged (Reissner-Nordstr\"om) BHs present two concentric horizons: the (outer) event horizon and the (inner) Cauchy horizon. In the limit of extremal charge, the horizons become degenerate. The cosmic censorship conjecture suggests that this extremal state cannot be exceeded via any finite physical processes. Yet extremally-charged BHs are of interest in their own right as they present intriguing features, such as: (i) zero surface gravity/Hawking temperature, (ii) a near-horizon instability \cite{Aretakis:2011hc, Lucietti:2012xr}, and (iii) equality between gravitational and electromagnetic absorption cross sections \cite{Oliveira:2011zz}.

The processes of absorption and scattering in the vicinity of black holes are potentially relevant for  experimental investigations. Since the 1960s, much theoretical work has been done on black hole scattering (cf., e.g., Ref.~\cite{Futt:1988} and references therein) in idealized scenarios. With the positive results of experiments performed at CERN in the search for the Higgs Boson \cite{Aad:2012tfa}, there is now an additional motivation for studying absorption and scattering of bosonic fields with mass. For example, it was recently suggested that accretion of dark matter onto compact objects will have a distinctive effect on extreme mass-ratio inspirals and their gravitational wave signatures \cite{Macedo:2013qea}.

The absorption of massive fields on the Schwarzschild spacetime was examined by Unruh \cite{Unruh:1976fm} nearly four decades ago. Scattering of massive fields by a Schwarzschild black hole was studied (for bosons and fermions) in \cite{Jung:2004yh,Doran:2005vm,Dolan:2006vj,Castineiras:2007ma}. The low-frequency absorption cross section for the charged massive scalar field in the $n$-dimensional Reissner-Nordstr\"om spacetime was analyzed in Ref.~\cite{Jung:2004yn}. Recent work on black holes and massive bosonic fields includes investigations of quasi-normal mode excitation \cite{Decanini:2014bwa,Decanini:2014kha}, long-lived modes in bosonic fields \cite{Rosa:2011my,Barranco:2012qs,Barranco:2013rua, Brito:2013xaa, Okawa:2014nda}, and superradiant instabilities \cite{Pani:2012vp, Dolan:2012yt, Yoshino:2013ofa}.

In this work we focus on the absorption cross section for a monochromatic planar wave of the neutral massive scalar field impinging upon a four-dimensional Reissner-Nordstr\"om spacetime. There are four parameters in this scenario: the mass $M$ and charge $Q$ of the BH, and the mass $\mass$ and frequency $\omega$ of the field. From these quantities, we may form three dimensionless parameters: the BH charge-to-mass ratio
$
q = |Q|/M  \label{qdef},
$
with $0 \le q < 1$ for sub-extremal BHs, and a pair of field-to-BH couplings, $M \omega$ and $M \mass$. Note that we adopt units in which $c= \hbar = G = 1$ so that, e.g., $M \mass \equiv M \mass / m_P^2$, where $m_P$ is the Planck mass. We also make use of an alternative dimensionless parameter,
\b
v = \sqrt{1 - \frac{\mass^2}{\omega^2}},  \label{vdef}
\e
corresponding to the ratio of the speed of propagation of the wave in the far-field to the speed of light. Here $0 < v \le 1$ for unbound modes, for which $\omega > \mass$. 

This paper is arranged as follows. In Sec.~\ref{sec:field} we review the theory of the scalar field in the Reissner-Nordstr\"om spacetime. In Sec.~\ref{sec:csec} we find the absorption cross section for the massive scalar field. In Secs.~\ref{sec:highfreq} and \ref{sec:lowfreq} we obtain the high- and low-frequency limits of the absorption cross section, respectively. In Sec.~\ref{sec:numerical} we present a selection of numerical results. We conclude with our final remarks and discussion in Sec.~\ref{sec:conclusion}. 

We adopt the spacetime signature $(+---)$ throughout.

\section{The Scalar Field\label{sec:field}}
The line element $ds^2 = g_{\mu \nu} dx^\mu dx^\nu$ of the Reissner-Nordstr\"om spacetime is
\b
ds^2 = f dt^2 - f^{-1}dr^2 - r^2d\theta^2 - r^2\sin^2{\theta} d\phi^2,
\label{le}
\e 
where
\b
f = \left(1-\frac{r_+}{r}\right)\left(1-\frac{r_-}{r}\right),
\e
and the horizon radii are
\b
r_{\pm} =M \pm \sqrt{M^2-Q^2} .
\e
The Klein-Gordon equation governing the propagation of a massive scalar field is 
\b
\nabla_\mu \nabla^\mu \Phi + m^2 \Phi = 0 ,
\label{kge}
\e
where $\nabla_{\mu}$ denotes the covariant derivative, and indices are raised with the inverse metric $g^{\mu \nu}$.

Without loss of generality, we will assume that the incoming wave is incident along the $z$-axis. An axially-symmetric solution to Eq.~(\ref{kge}) in Reissner-Nordstr\"om spacetime can be written as
\b
\Phi_{\omega l} = \frac{\psi_{\omega l}(r)}{r} P_l(\cos \theta) e^{-i\omega t},\hspace*{0.2in} \omega > m ,
\label{eq:separation}
\e
where $P_l(\cos \theta)$ is a Legendre polynomial, and $\psi_{\omega l}(r)$ satisfies the radial equation
\b
\frac{d^2}{dr^2_*}\psi_{\omega l} +\left[\omega^2 -\Veff(r)\right]\psi_{\omega l} = 0,
\label{fdr}
\e
with the effective potential
\b
\Veff(r) = f\left(\mass^2+\frac{l(l+1)}{r^2}+\frac{2M}{r^3}-\frac{2Q^2}{r^4}\right) .
\label{pte}
\e
Here we have used the tortoise coordinate $r_*$, defined in the standard way by
$
dr_* / dr = f^{-1} 
\label{trc}
$.

\begin{figure*}
\includegraphics[width=\columnwidth]{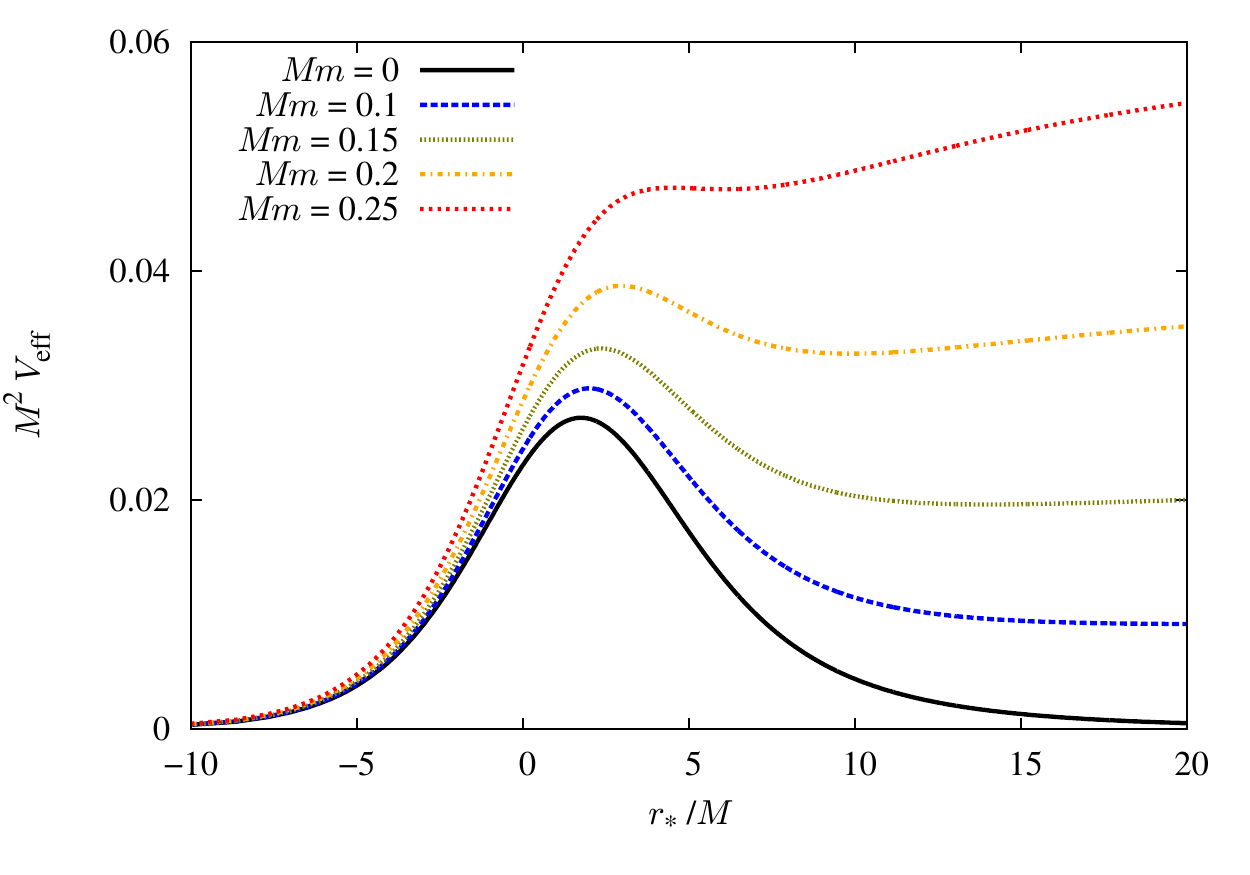}
\includegraphics[width=\columnwidth]{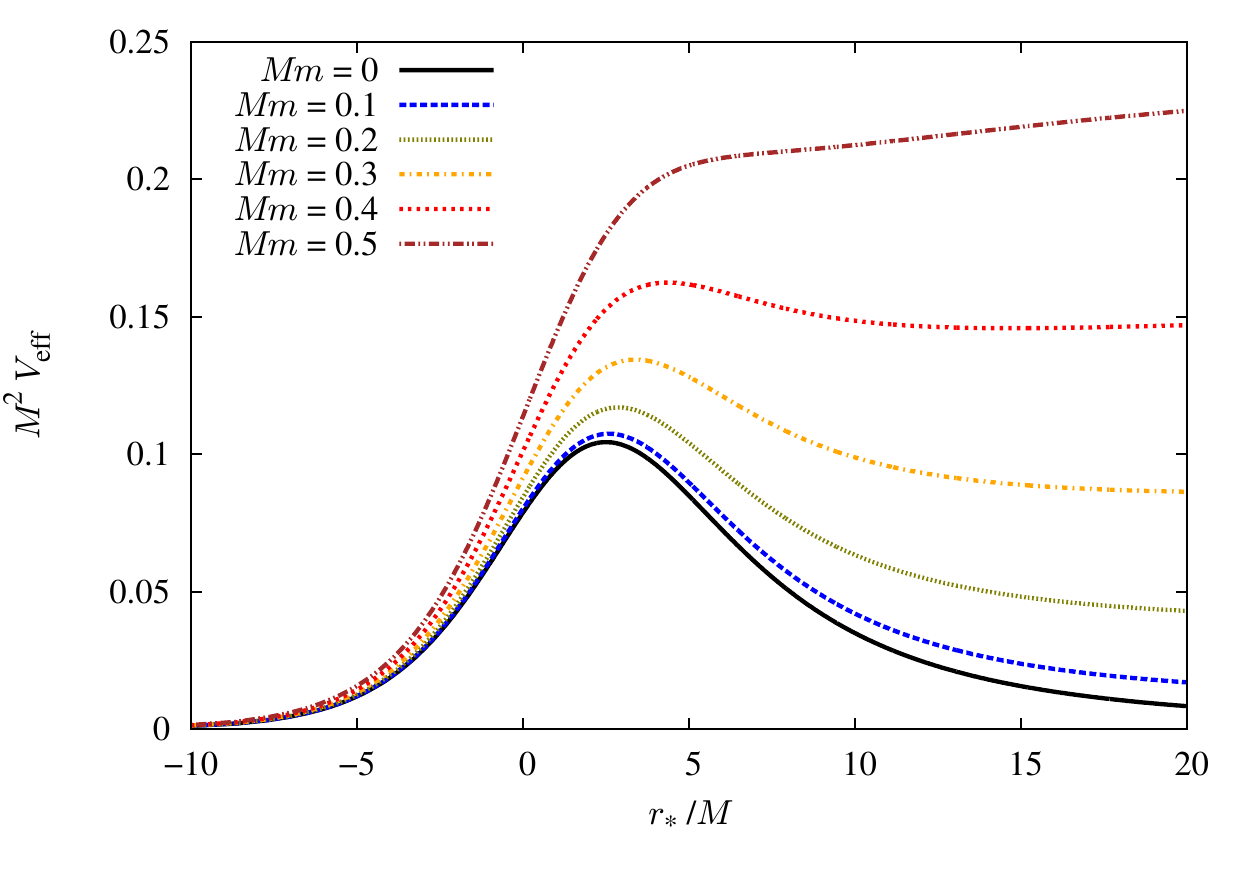}
\caption{Effective potential for $q\equiv Q/M=0.4$, $l=0$ (left, $\Mm_c=0.195$) and $l=1$ (right, $\Mm_c=0.405$) plotted for different values of the scalar field mass.}
\label{fig:effpot}
\end{figure*}

Equation~(\ref{fdr}) is a Schr\"odinger-like equation with an effective potential. Figure \ref{fig:effpot} shows $\Veff$, defined in Eq.~(\ref{pte}), for $l=0, 1$ and various values of the scalar field mass. In the far field, the potential may be expanded as $\Veff = m^2 - 2M m^2 / r + [l (l+1) + Q^2 m^2]/r^2 + O(r^{-3})$. The mass coupling term generates a Newtonian-like attraction at $\mathcal{O}(r^{-1})$, and the angular momentum $l$ (and charge $Q$) creates a potential barrier at $\mathcal{O}(r^{-2})$. In the limit $r \rightarrow r_+$, the potential tends to zero.  Figure \ref{fig:effpot} shows that, for moderate values of $M\mass$, the effective potential admits a local maximum and local minimum. These features are washed out as $\Mm$ increases.

\begin{figure}
\includegraphics[width=\columnwidth]{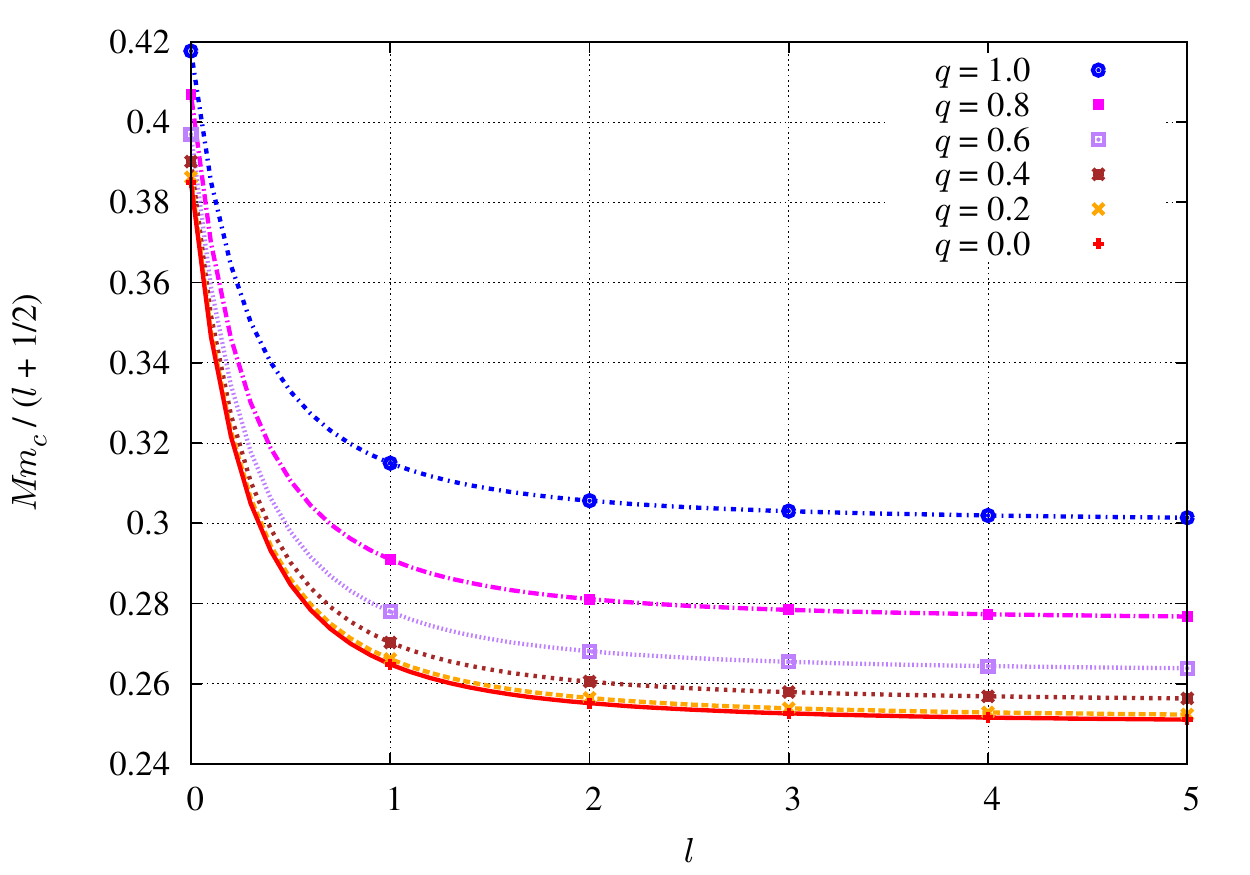}
\caption{Critical mass coupling $Mm_c$ \cite{Jung:2004yh} as a function of multipole $l$. For $l=0$, $\Mm_c \approx 0.192$ in the Schwarzschild ($q=0$) case, and $\Mm_c \approx 0.209$ in the extremal Reissner-Nordstr\"om ($q = 1$) case. In the large-$l$ regime, $Mm_c(q = 0,l) \approx 0.25 (l + 1/2) $ and $Mm_c(q=1, l) \approx 0.3 (l + 1/2)$.}
\label{fig:critmass}
\end{figure}

Jung and Park \cite{Jung:2004yh} introduced the notion of a `critical mass' $\Mm_c$, defined (for each $l$ and $q$) as the value at which the local maximum value of $\Veff$ is equal to the asymptotic value, $\Veff(r\rightarrow \infty) = m^2$. For $M m > \Mm_c$, all unbound modes are strongly absorbed, regardless of mode frequency. In the large-$l$ regime, the critical mass scales linearly with $l + 1/2$. Figure \ref{fig:critmass} shows $\Mm_c / (l+1/2)$ as function of $l$, determined numerically, for various black hole charge-to-mass ratios $q$. We see that $\Mm_c$ increases somewhat with $q$.

Let us now turn our attention to the asymptotic solutions of Eq.~(\ref{fdr}). Since we are interested in the absorption process, we consider only those modes which are ingoing at the outer horizon,
\b
\psi_{\omega l}(r) \approx 
\left\{ 
\begin{array}{ll}
\sqrt{v} \, T_{\omega l} \, e^{-i\omega r_*}, \quad &\mbox{for $r\rightarrow r_+$},\\
 e^{-i\varrho} + R_{\omega l} e^{i\varrho}, \quad &\mbox{for $r\rightarrow \infty$},
\end{array}
\right.
\label{sol}
\e
where $T_{\omega l}$, $R_{\omega l}$ are complex coefficients, $v$ was defined in Eq.~(\ref{vdef}), and $\varrho = \varrho(r)$ has the leading-order expansion
\b
\varrho = \omega v r + \frac{\omega M (1+v^2)}{v} \ln(2 M \omega v r) + \mathcal{O}(r^{0}).  \label{vartheta-def}
\e

Here we note that the normalization of $\psi_{\omega l}(r)$ has been chosen for later convenience. $|R_{\omega l}|^2$ and $|T_{\omega l}|^2$ may be interpreted as reflection and transmission coefficients, respectively. By considering the Wronskian of Eq.~(\ref{fdr}), it is straightforward to show that 
\b
|R_{\omega l}|^2 + |T_{\omega l}|^2 = 1,
\label{rtf}
\e
representing the conservation of flux (cf. Sec.~\ref{sec:csec}).

\section{Absorption Cross Section\label{sec:csec}}
In this section we obtain an expression for the absorption cross section as a sum of partial wave contributions. 
We seek a field $\Phi$ which is purely ingoing at the event horizon [cf.~Eq.~(\ref{sol})] and which, in the far-field, resembles the sum of an incident planar wave $\Phi^I$ and an outgoing scattered wave $\Phi^S$. The absorption cross section is defined as the ratio of the flux in $\Phi$ passing into the black hole, to the current in the incident wave $\Phi^I$.

We take the incident wave $\Phi^I$ to be a monochromatic planar wave of frequency $\omega$ which, without loss of generality, we assume to be propagating along the $z$-axis. In a Minkowski spacetime, one may write $\Phi^I_{\text{(M)}} = e^{-i\omega (t - v z)}$, and then make use of 
\b
e^{i \omega v z} = \sum_{l=0}^\infty (2 l + 1) i^l j_l(\omega v r) P_l(\cos \theta),  \label{PhiI}
\e
to expand in partial waves. Here $j_l(\cdot)$ is a spherical Bessel function. In the far-field, this becomes
\b
\Phi^I_{\text{(M)}} \sim \frac{e^{-i \omega t}}{r} \sum_{l = 0}^\infty c_{l\omega} \left(e^{-i \omega v r} + e^{-i \pi (l+1)} e^{i \omega v r}\right) P_l(\cos \theta),
\e
where  
\b
c_{l\omega} = \frac{2 l + 1}{2 i \omega v} e^{i \pi (l+1)} .
\e
By contrast, in a black hole spacetime the long-ranged nature of the gravitational field means that a planar wave is {\it distorted}, even far from the black hole. Taking note of Eqs. (\ref{sol}) and (\ref{vartheta-def}), the analogue of a planar wave has an asymptotic form
\b
\Phi^I \sim \frac{e^{-i \omega t}}{r} \sum_{l = 0}^\infty c_{l\omega} \left(e^{-i \varrho} + e^{-i \pi (l+1)} e^{i \varrho}\right) P_l(\cos \theta).
\e
The physical solution $\Phi$ is constructed from the horizon-ingoing modes (\ref{sol}) in such a way that, in the far-field, the ingoing part of $\Phi$ matches on to the ingoing part of $\Phi^I$. That is, we define
\b
\Phi = \frac{e^{-i \omega t}}{r} \sum_{l=0}^\infty c_{l \omega} \psi_{\omega l}(r) P_l(\cos \theta) .
\label{phi-def}
\e
The scattered wave $\Phi^S = \Phi - \Phi^I$ has the asymptotic form
\b
\Phi^S \sim \frac{e^{-i \omega t}}{r} \hat{f}(\theta) e^{i \varrho},
\e
with a scattering amplitude $\hat{f}(\theta)$ given by
\b
\hat{f}(\theta) = \frac{1}{2i \omega v} \sum_{l=0}^\infty (2l+1) \left( e^{i\pi(l+1)} R_{\omega l} - 1 \right)  P_l(\cos \theta) .
\e

To find the absorption cross section, we may begin by introducing a four-current
\b
J_\alpha =\frac{i}{2}\left[\Phi^* \nabla_\alpha \Phi - \Phi \nabla_\alpha \Phi^*\right],
\label{jto}
\e
which satisfies the conservation law $\nabla_\alpha J^\alpha = 0$ by Eq. (\ref{kge}). Now, we consider a four-volume bounded by 3-surfaces defined by $t=t_1$, $t = t_2$, $r=r_1$ and $r=r_2$ (where $t_1 < t_2$ and $r_+ < r_1 < r_2$). Applying Gauss' theorem and taking the limit $t_2 - t_1 \rightarrow 0^+$, leads to
\b
\frac{d}{dt} \left\{ \int r^2 J^t dr_\ast d\Omega \right\}  = \left[ \phantom{\frac{}{}}  N(r)  \phantom{\frac{}{}}   \right]^{r_2}_{r_1} .
\label{conslaw}
\e

Here, $N(r)$ is the flux passing through a surface of constant radius $r$, given by
\b
N(r) = - \int r^2 \, J^r d\Omega  .
\e
We consider a stationary scenario, in which the left-hand side of Eq. (\ref{conslaw}) is zero, and thus $N(r_1) = N(r_2) = N$. In this case, $N$ is (minus) the flux of particles passing into the black hole \cite{Unruh:1976fm}. 

The absorption cross section $\sigma$ is defined as the ratio of $|N|$ to the incident current in the planar wave, $\omega v$. We may insert Eq. (\ref{phi-def}) into Eq. (\ref{jto}) and use the orthogonality of Legendre polynomials [$\int P_l(\cos \theta) P_{l'} (\cos \theta) d\Omega = 4\pi \delta_{ll'} / (2l+1)$] to write the total absorption cross section $\sigma$ as a sum of partial cross sections $\sigma_l$,
\b
\sigma = \sum_{l=0}^\infty \sigma_l ,  \label{acs}
\e
defined in terms of modal transmission/reflection coefficients by
\b
\sigma_l = \frac{\pi (2l+1)}{\omega^2 v^2} \left| T_{\omega l} \right|^2 = \frac{\pi (2l+1)}{\omega^2 v^2} \left(1 -  \left| R_{\omega l} \right|^2 \right).  \label{pcs}
\e

\section{High-frequency regime\label{sec:highfreq}}

In the limit of high frequency, the wavelength of the field becomes very small in comparison to the scale of the black hole (e.g.~the horizon radius). Under the eikonal approximation, a wavefront propagates along geodesics of the spacetime \citep{Futt:1988}. The `geodesic capture cross section' is defined as $\sighf = \pi b_c^2$, where $b_c$ is the critical impact parameter corresponding to the unbound geodesic which asymptotically approaches the unstable circular orbit at $r=r_c$. The critical impact parameter $b_c$ may be found by solving the orbital equation for a timelike geodesic in the Reissner-Nordstr\"om spacetime. Without loss of generality, we may consider motion in the equatorial plane ($\theta = \pi/2$).  Let us start from the `energy' equation,
\b
\dot{r}^2 \equiv \mathcal{T}(r) = \mathcal{E}^2 - f (m^2 + \mathcal{L}^2 / r^2)  \label{eq:energy}
\e
where $\dot{r} = dr/d\tau$, $\tau$ is the proper time, and $\mathcal{E} =  f \dot{t}$ and $\mathcal{L} = r^2 \dot{\phi}$ are the energy and angular momentum of the geodesic, respectively. Now we introduce the impact parameter $b \equiv \mathcal{L} / (\mathcal{E} v)$, where $v^2 = 1 - m^2 / \mathcal{E}^2$. (We note in passing that $v$ defined above [for a geodesic] is equivalent to $v$ defined in Eq.~(\ref{vdef}) [for a field/wave] under the standard semi-classical mapping $\mathcal{E} \leftrightarrow \omega$, $\mathcal{L} \leftrightarrow l + 1/2$.) This allows us to write 
\b
\frac{\mathcal{T}}{\mathcal{L}^2} = \frac{1}{b^2 v^2} - f \left( \frac{1-v^2}{b^2 v^2} + \frac{1}{r^2} \right).
\label{phf}
\e 
To obtain $b_c$, the critical impact parameter, and $r_c$, the radius of the unstable circular orbit (or `critical radius'), we set this equation and its radial derivative to zero, i.e.,~$\mathcal{T} = 0$ and $d\mathcal{T} / dr = 0$. This yields
\b
b_c= \frac{r_c}{v f_c^{1/2}}\left[1 - (1-v^2)f_c  \right]^{1/2},  \label{bc-def}
\e
where $f_c = f(r_c)$, and a quartic equation for $r_c$, namely,
\begin{widetext}
\b
r^4_c + M\frac{(1-4v^2)}{v^2}r^3_c +
\left(\frac{4(v^2-1)}{v^2}M^2+2Q^2\right)r^2_c + \frac{4MQ^2(1-v^2)}{v^2}r_c +\frac{Q^4(v^2-1)}{v^2} = 0.
\label{rc}
\e
\end{widetext}

\begin{figure}[h]
\includegraphics[width=\columnwidth]{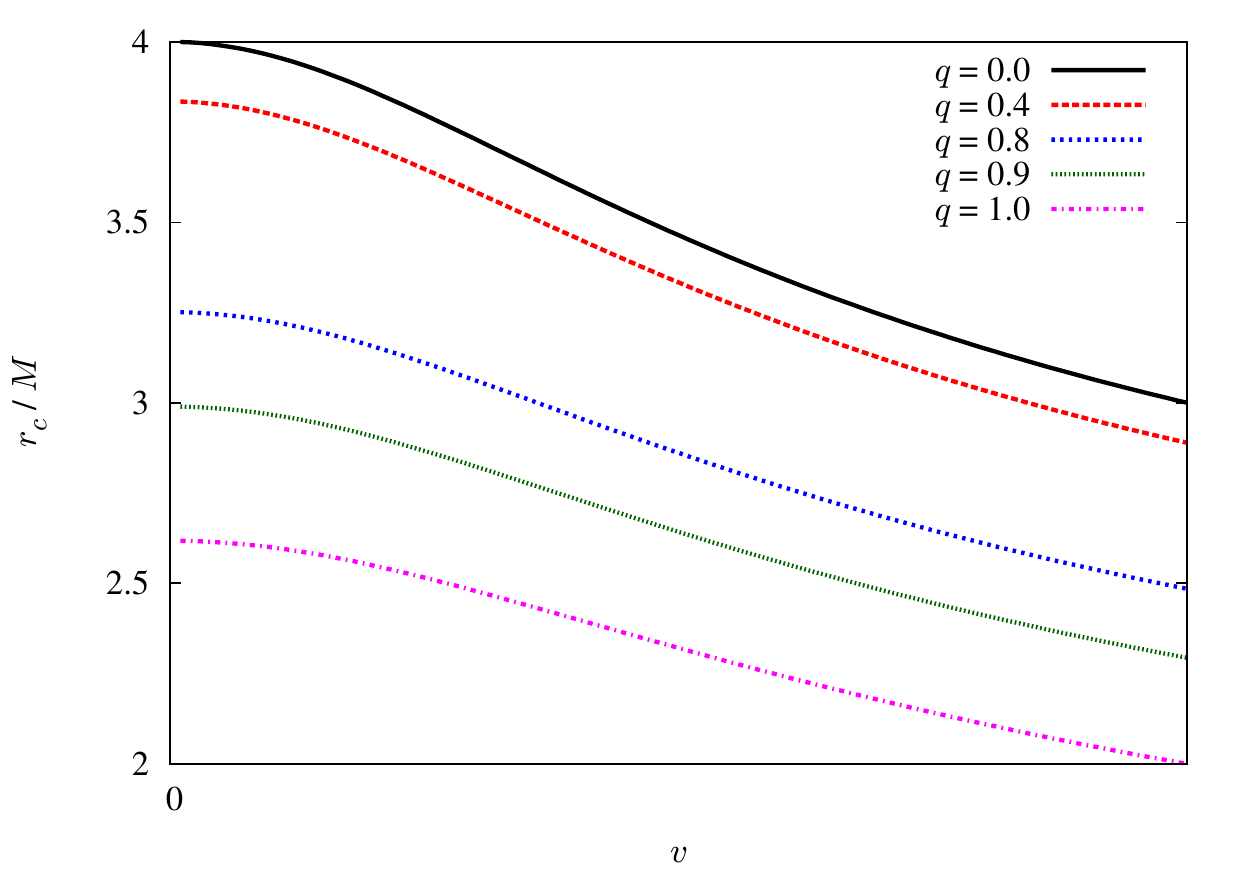}
\includegraphics[width=\columnwidth]{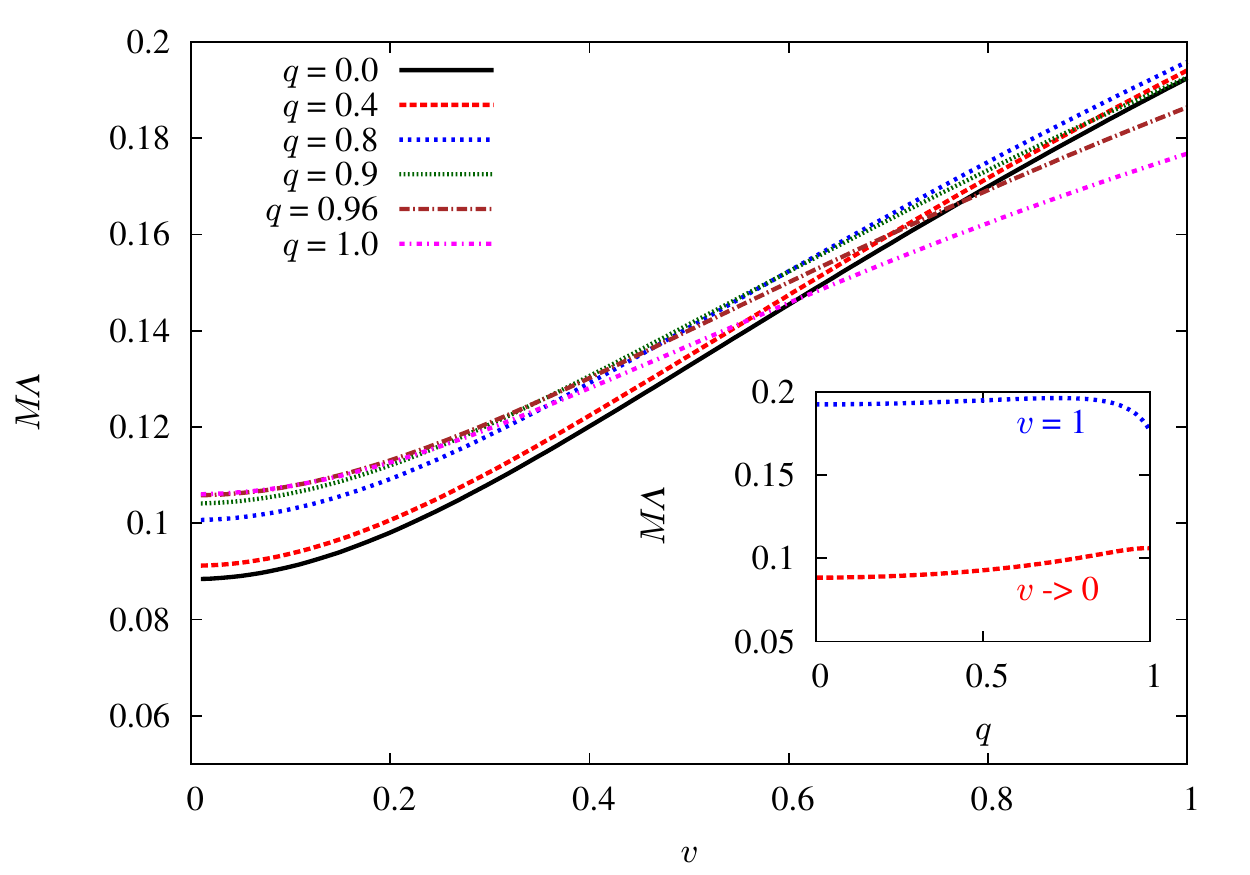}
\caption{Orbital parameters for critical geodesics on Reissner-Nordstr\"om spacetime. The top plot shows the radius of the unstable circular orbit, $r_c$, as a function of incident speed $v$ for a range of charge ratios $q$. The bottom plot shows the Lyapunov exponent $\Lambda$ associated with the peak in the potential barrier. The inset shows the Lyapunov exponent as a function of $q$ in the nearly-bound ($v\rightarrow 0$) and null ($v=1$) cases.
The critical impact parameter $b_c$ may be inferred from the absorption function shown in Fig.~\ref{fig:absfn}. }
\label{fig:orbparams}
\end{figure}

We seek a root of Eq.~(\ref{rc}) that is larger than the outer horizon, i.e., $r_c > r_+$, and which corresponds to a local minimum of the right-hand side of Eq.~(\ref{phf}). This root may be found numerically. The top plot of Fig.~\ref{fig:orbparams} shows the critical radius as a function of $v$ for various charge ratios $q$. We see that, in general, $r_c$ decreases as $v$ increases, and as $q$ increases.

In the limit $v \rightarrow 0$, the critical impact parameter $b_c$ diverges as $1/v$. Let us therefore introduce a dimensionless `absorption function' $\FF(v,q) = v^2 b_c^2 / M^2$, which is regular in this limit, so that in the high-frequency regime
\b
\sigma \rightarrow \sighf = \FF(v,q) \, \frac{\pi M^2}{v^2}. \label{Fdef} 
\e
In the null geodesic case ($v = 1$) \cite{Crispino:2008zz},
\b
\FF(1, q) = \frac{\left( 3 + \sqrt{9 - q^2} \right)^4}{8\left(3 - 2q^2 + \sqrt{9 - 8q^2}\right)} . \label{eq:F1q}
\e
In the Schwarzschild case ($q = 0$) \cite{Dolan:2006vj}, 
\b
\FF(v,0) = \frac{1}{4}\frac{(4v^2-1+\sqrt{1+8v^2})^2}{\sqrt{1+8v^2} - 1} \left( 3+\sqrt{1+8v^2} \right) .
\e
In the extremal case ($q = 1$),
\b
\FF(v,1) =  \chi^2 \frac{v^2+\frac{2(1-v^2)}{\chi}-\frac{(1-v^2)}{\chi^2}}{1-\frac{2}{\chi}+\frac{1}{\chi^2}},\label{eq:Fv1}
\e
where
\b
\chi = \frac{1}{6}\frac{\kappa^{1/3}}{v^2}+ \frac{2}{3}\frac{(3v^2+1)}{v^2\kappa^{1/3}}+\frac{1}{3}\frac{(3v^2-1)}{v^2}
\e
and
\b
\kappa=-36v^2+108v^4-8+12\sqrt{(27v^4-22v^2-5)3} \, v^2.
\e
In the limit $v \rightarrow 0$, we may find $r_c$ by solving the cubic
\b
x^3 - 4x^2 + 4q^2 x - q^4 = 0,
\e
where $x  = r_c/M$, and substituting the solution into Eq. (\ref{bc-def}) to obtain $b_c$ and thus $F(0, q)$.

\begin{figure}[h]
\includegraphics[width=\columnwidth]{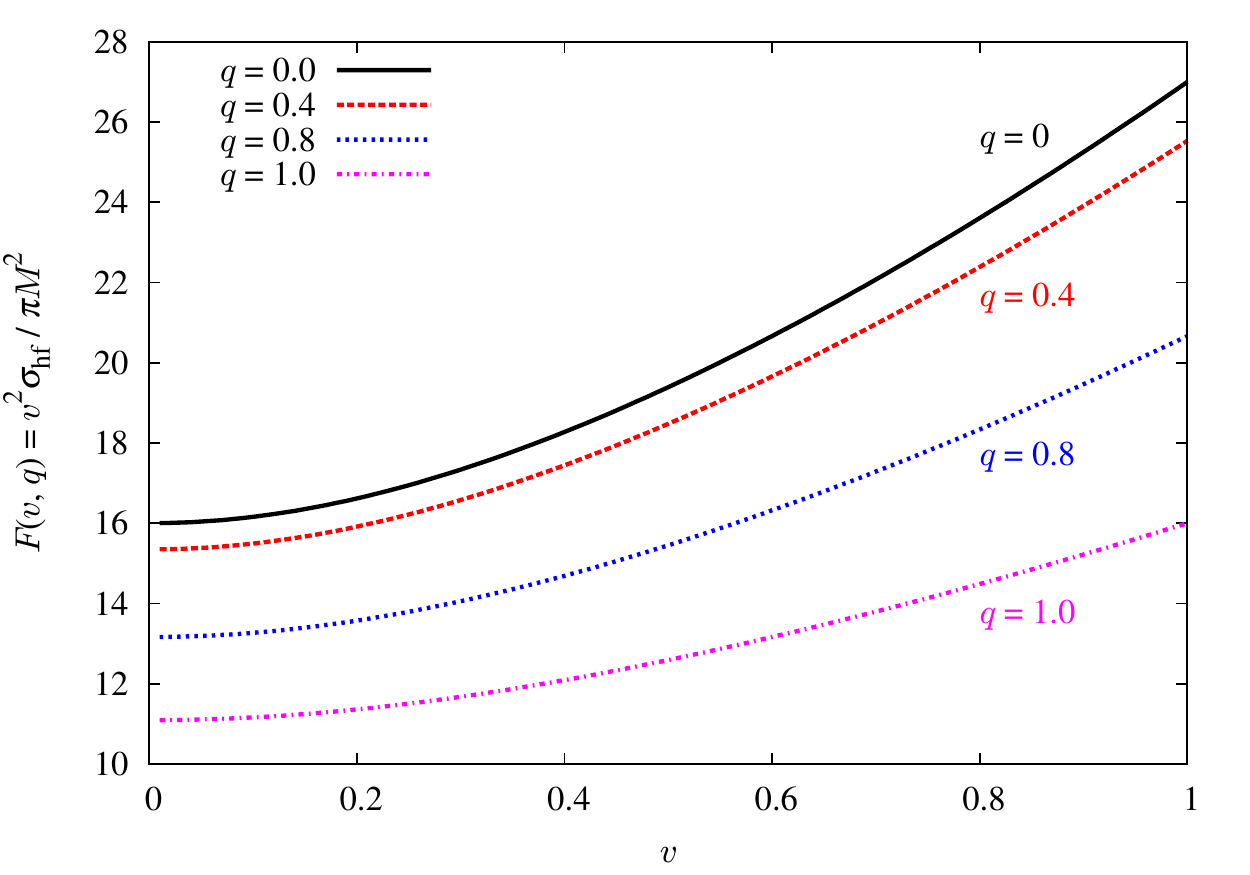}
\caption{$F(v,q)$ as a function of the incident speed $v$ for various charge-to-mass ratios $q$. $F(v,q)$ determines the high-frequency absorption cross section via Eq.~(\ref{Fdef}). For closed-form solutions in limits $v\in\{0,1\}$, $q \in \{0, 1\}$ see Eqs.~(\ref{eq:F1q})--(\ref{eq:Fv1}).}
\label{fig:absfn}
\end{figure}

In the general case ($v,q \neq 0,1$), one may compute the values of the critical ray by finding the numerical solution of Eq.~(\ref{rc}). We find that the absorption function $F(v, q)$ varies smoothly, as shown in Fig.~\ref{fig:absfn}. In Fig.~\ref{fig:tacsm} we compare the values from the geodesic analysis with the total absorption cross section at moderate-to-large frequencies $M\omega$. 

\subsection{Sinc approximation\label{sec:sinc}}
In the high-frequency regime the absorption cross section exhibits regular oscillations (with $\omega$) around the limiting value (see e.g.~Fig.~\ref{fig:tacsm}). In the case of a massless scalar field absorbed by a Schwarzschild BH, Sanchez \cite{Sanchez:1978} found that a simple formula provided a good fit at high frequencies, 
\b
\sighf^{(q=0,v=1)} \approx 27 \pi M^2 \left[ 1 - \hat{\alpha} \,  \sinc(2\pi \sqrt{27} M\omega) \right],
\e
where $\sinc(x)= \sin(x) / x$ and $\hat{\alpha} \approx \sqrt{32/27}$ (see Eq.~(30) in Ref.~\cite{Sanchez:1978}).  Decanini, Folacci and coworkers~\cite{Decanini:2011xi, Decanini:2011xw} have applied the complex angular momentum approach to analyze the absorption cross section. They derived Sanchez' result in the high-frequency regime, with a more accurate coefficient of $\hat{\alpha} = 8 \pi e^{-\pi}$, and gave higher-order corrections. Furthermore, Ref.~\cite{Decanini:2011xi} showed that regular oscillations are a universal feature of cross sections for massless fields absorbed by spherically-symmetric BHs. 

We now seek to extend the complex angular momentum analysis to the massive-field case (see also Ref.~\cite{Decanini:2011eh}). As before, the oscillatory contribution to the cross section is related to a sum of the residues of so-called Regge poles, and the asymptotic properties of the Regge pole spectrum may be determined by geodesic analysis. We used the approach of Ref.~\cite{Dolan:2009nk} to show that, in the high-frequency regime, the Regge pole $\lambda_n$ is approximately
\b
\lambda_n = v b_c \omega + i (n+1/2) \hat{\beta} + \mathcal{O}(\omega^{-1}) ,
\e
where $\hat{\beta} = v b_c \Lambda$ and $\Lambda$ is the Lyapunov exponent associated with the unstable circular orbit, i.e.,
\b
\Lambda = \frac{1}{\dot{t}} \sqrt{\frac{1}{2} \frac{d^2 V_r}{dr^2}} = \frac{v f_c }{r_c} \sqrt{k_c}, \label{eq:lyapunov}
\e
where
\begin{eqnarray}
k_c &=& \frac{1}{v^2 r_c^4 f_c} \left\{(4v^2 - 1)Mr_c^3 + [8M^2(1-v^2) - 4Q^2v^2] r_c^2 \right. \nonumber \\ 
&& \quad \left. - 12MQ^2(1-v^2) r_c + 4Q^4 (1-v^2) \right\} .
\end{eqnarray}
The Lyapunov exponent is shown in the lower plot of Fig.~\ref{fig:orbparams}.

Following the steps in Ref.~\cite{Decanini:2011xi}, it is possible to show that the high-frequency approximation to the cross section is given by
\b
\sighf \approx \frac{F(v,q) \pi M^2}{v^2} \left[ 1 - 8 \pi \hat{\beta} e^{-\pi \hat{\beta}} \, \sinc\left(2 \pi v b_c \omega\right) \right] . 
\label{eq:sinc}
\e
This approximation is compared with numerically-determined cross sections in Sec.~\ref{sec:numerical}, showing excellent agreement.

\section{Low-frequency regime\label{sec:lowfreq}}

In this section we analyze the low-frequency limit of the Reissner-Nordstr\"om absorption cross section for the massive scalar field, following the method of Ref.~\cite{Unruh:1976fm}. Note that, since $\omega > m$, the low-frequency regime($M \omega \ll 1$ implies that $Mm \ll 1$.

We will first analyze the case for general Reissner-Nordstr\"om black holes $(r_+>r_->0)$ and then specialize to the cases of the Schwarzschild ($q=0$, $r_+=2M$, $r_-=0$) and extreme Reissner-Nordstr\"om black holes ($q=1$, $r_+=r_-=M$).

We consider three different regions: the region very close to the black hole (region \textit{I}), an intermediate region, in which the frequency and mass terms are much smaller than the other contributions in Eq.~(\ref{fdr}) (region \textit{II}), and a region distant from the black hole (region \textit{III}). We match together the solutions so obtained to get a global solution. 

\subsection{General case}

We may rewrite the differential equation (\ref{fdr}) as
\b
\frac{f}{r^2}\frac{d}{dr}\left(fr^2\frac{d}{dr}\varphi\right)+(\omega^2 - V_{RN}(r))\varphi = 0,
\label{fdr2}
\e
where $\varphi=r^{-1} \psi_{\omega l}$ [cf.~Eq.~(\ref{eq:separation})] and
\b
V_{RN}= f \left(m^2 + \frac{l(l+1)}{r^2}\right).
\e
For region \textit{I} $(r \approx r_+)$, Eq.~(\ref{fdr2}) is approximately
\b
\frac{d^2\varphi}{dr_*^2} + \omega^2 \varphi = 0,
\label{ert}
\e
with $\varphi^I_{RN} \propto e^{-i\omega r_*}$
representing a transmitted wave. We may write the tortoise coordinate explicitly as a function of $r$, as
\b
r_* = r + \frac{r_+^2}{r_+ - r_-}\ln(r-r_+)-\frac{r_-^2}{r_+ - r_-}\ln(r-r_-),
\label{torn}
\e
after fixing the constant of integration appropriately. Let us consider the dominant term of Eq.~(\ref{torn}), for $r\rightarrow r_+$, 
\b
r_* \sim \frac{r_+^2}{r_+ - r_-}\ln(r-r_+) + r^{(0)}_* ,
\e
where $r^{(0)}_*$ is a constant, so that
\b
\varphi_{RN}^I = \ARNtra \left| r - r_+ \right|^{-i\omega \alpha} .
\label{rn1}
\e
Here $\ARNtra$ is a complex constant, and $\alpha = r_+^2 / (r_+ - r_-)$.

In order to find the solution in region \textit{II} we take the limit $\omega\rightarrow 0$, $m \rightarrow 0$ in Eq.~(\ref{fdr2}). Since we are interested in computing the absorption cross section in the limit $M \omega , \Mm \ll 1$, we may restrict ourselves to the $l=0$ mode, which is the dominant term in this limit \footnote{In Ref. \cite{Unruh:1976fm} it is shown that the low-frequency limit of the transmission coefficient for small black holes behaves as $\omega^{2l+2}$, so that the $l=0$ contribution is dominant in this limit.}. Thus, the differential equation reduces to
\b
\frac{d^2}{dr^2}\varphi_{RN} - \frac{(r_+ + r_- - 2r)}{(r-r_+)(r-r_-)}\frac{d}{dr}\varphi_{RN} = 0 ,
\label{er2}
\e
with solution given by
\b
\varphi_{RN}^{II} = \zeta \ln\left(\frac{r-r_+}{r-r_-}\right) + \tau.
\label{rn2}
\e
where $\zeta$ and $\tau$ are constants to be determined. 

We now seek an overlap between the solutions in regions $I$ and $II$. We may rewrite Eq.~(\ref{rn1}) as
\b
\varphi_{RN}^I \approx \ARNtra \left(1-i\omega \alpha \ln(r-r_+)\right).
\label{rn12}
\e
If we take the limit $r \rightarrow r_+$ in Eq.~(\ref{rn2}) we obtain
\b
\varphi_{RN}^{II} = \zeta \ln(r-r_+)-\zeta \ln(r_+-r_-) + \tau.
\label{rn21}
\e
Comparing Eqs.~(\ref{rn12}) and (\ref{rn21}) yields
\b
\begin{array}{ll}
\zeta = -i\omega \alpha \ARNtra,\\
\tau = \left(1-i\omega \beta\right) \ARNtra,
\end{array}
\label{coen1}
\e
where $\beta = \alpha \ln(r_+-r_-)$.

For region \textit{III} ($r \gg r_+$) we can rewrite Eq. (\ref{fdr2}) as
\begin{eqnarray}
\left\{ \frac{d^2}{dr^2} \right. &+& \left[(\omega^2-m^2)+\frac{2M(2\omega^2 - m^2)}{r} \right. \nonumber \\ 
 && \left. \left. -\frac{l(l+1)}{r^2}\right]\right\}r f^{1/2}\varphi =0,
\end{eqnarray}
where we neglect terms of $\mathcal{O}\left(1/r^2\right)$ that are proportional to $\omega^2$ and $m^2$, and terms of order $1 / r^3$ and higher. The solution to the above equation can be written as:
\b
\varphi^{III}_{RN} = a\frac{F_l(\eta, \omega vr)}{r}+ b\frac{G_l(\eta,\omega vr)}{r},
\label{sr3}
\e
where $\eta = -M\omega(1+v^2)/v$, and $F_l(\eta,x)$ and $G_l(\eta,x)$ are the regular and irregular Coulomb wave functions, respectively \cite{Abram:1972}. In the far-field, we may write
\b
\varphi^{III}_{RN} \approx \ARNref e^{i\vartheta} +  \ARNinc e^{-i\vartheta}
\e
where $\vartheta = \omega v r - l\pi/2 - \eta \ln (2 M \omega v r) + \text{arg}\Gamma( l + 1 + i \eta )$. 
Here $\ARNinc$ and $\ARNref$ are related to $a$ and $b$ by
\b
\ARNinc =  \frac{-a+ib}{2i} ,\hspace{0.2in} \ARNref =  \frac{a+ib}{2i}  .
\label{aba}
\e
For $\omega r\ll 1$ and $l=0$, Eq.~(\ref{sr3}) reduces to
\b
\varphi_{RN}^{III} = a\rho \omega v + \frac{b}{\rho r},
\label{s32}
\e
where we have used for the Coulomb wave functions
\b
F_0(\eta,x) = \rho x, \hspace{0.2in} G_0(\eta,x) = \frac{1}{\rho},
\e
and
\b
\rho^2 = \frac{\eta}{e^{\eta} - 1}=\frac{-M\omega(1+v^2)/v}{e^{-M\omega(1+v^2)/v} - 1}.
\e
In the asymptotic limit, Eq.~(\ref{rn2}) becomes
\b
\varphi^{II}_{RN} = -\zeta\frac{r_+ - r_-}{r} + \tau.
\label{rn23}
\e
Using Eqs. (\ref{coen1}), (\ref{s32}) and (\ref{rn23}), we find
\b
\begin{array}{ll}
a=\frac{\ARNtra}{\rho \omega v}\left(1-i\omega\beta\right),\\
b=ir_+^2\omega \rho \ARNtra.
\end{array}
\label{crn2}
\e

We substitute Eq.~(\ref{crn2}) in Eq.~(\ref{aba}) and obtain
\b
\begin{array}{ll}
\ARNinc = -\ARNtra \frac{\left(1+r_+^2\omega^2 \rho^2 v-i\omega\beta\right)}{2i\rho\omega v},\\
\ARNref =\ARNtra \frac{\left(1-r_+^2\omega^2 \rho^2 v-i\omega \beta\right)}{2i\rho \omega v}.
\end{array}
\label{con3}
\e
The reflection coefficient is given by
\b
\left|R_{\omega l}\right|^2 = \left|\frac{\ARNref}{\ARNinc}\right|^2 = \left| \frac{1-r_+^2\omega^2 \rho^2 v -i\omega \beta}{1+r_+^2\omega^2 \rho^2 v-i\omega\beta } \right|^2,
\e
which, recalling Eq. (\ref{pcs}), gives for the absorption cross section, for $l=0$, in the approximation $\omega \approx 0$ and $m \approx 0$:
\b
\siglf =  \frac{\pi}{\omega^2 v^2}\left(\frac{4r_+^2\omega^2 \rho^2 v}{\left(1+r_+^2\omega^2 \rho^2 v \right)^2 + \omega^2 \beta^2}\right).
\label{lfcn}
\e
In the low-frequency limit we can also consider $\rho\approx 1$ and take only the first term in the denominator of Eq. (\ref{lfcn}), so that we are left with
\b
\siglf = \frac{\mathcal{A}}{v},
\label{lfcn2}
\e
where $\mathcal{A} = 4\pi r_+^2$ is the area of the Reissner-Nordstr\"om black hole.

In Sec.~\ref{sec:numerical} we compare this low-frequency limit with the numerical results.

\subsection{Schwarzschild case}

Having outlined the procedure for the Reissner-Nordstr\"om case in the previous section, we can directly find the results for the Schwarzschild case by inserting $r_+ = 2M$ and $r_-=0$ in the previous expressions. We have then, from Eq. (\ref{lfcn}),

\b
\siglf^{\text{Schw}} =  \frac{\pi}{\omega^2 v^2}\left(\frac{16\omega^2 M^2\rho^2 v}{\left(1+4\omega^2 \rho^2 v \right)^2+4\omega^2 M^2\ln^2(2M)}\right).
\label{lfc}
\e
Considering $\rho\approx 1$ and taking only the first term in the denominator of Eq. (\ref{lfc}) leads to 
$
\siglf^{\text{Schw}} = 16\pi M^2 / v .
$
This result was originally obtained by Unruh \cite{Unruh:1976fm}.

\subsection{Extreme case}
For the extreme Reissner-Nordstr\"om case, in which $r_+ = r_-$, we can repeat the argument with minor modifications. We have Eq. (\ref{fdr2}) with $V_{RN} \rightarrow V_e$, where
\b
V_e=\left(1-\frac{M}{r}\right)^2\left(m^2 + \frac{l(l+1)}{r^2}\right).
\e

The tortoise coordinate for the extreme case is
\b
r_* = r + 2M\ln(r-M)-\frac{M^2}{r-M}.
\label{tore}
\e
We consider only the dominant term of Eq. (\ref{tore}), for $r\rightarrow M$, i.e.,
\b
r_* \sim -\frac{M^2}{r-M} + r_*^{(0)} .
\e
The solution in region $I$ is given by
\b
\varphi_e^I = A^{\text{tra}}_e \exp\left( i \omega M^2 / (r-M) \right).
\label{rr1}
\e
For region \textit{II} the radial equation reduces to
\b
\frac{d^2}{dr^2}\varphi_e + \frac{2}{(r-M)}\frac{d}{dr}\varphi_e = 0,
\e
with solution given by
\b
\varphi_e^{II} = \frac{\zeta_e}{r-M} +  \tau_e.
\label{rr2}
\e
If we take the limit $\omega \rightarrow 0$ in Eq. (\ref{rr1}), we obtain
\b
\varphi_e^I = \Aetra\left(1+i\frac{\omega M^2}{r-M}\right).
\label{r12}
\e
Comparing Eqs. (\ref{rr2}) and (\ref{r12}) yields
\b
\begin{array}{ll}
\zeta_e = i\omega M^2 \Aetra,\\
\tau_e = \Aetra.
\end{array}
\label{coe1}
\e

For region \textit{III} we have again the solution (\ref{sr3}) and for low frequencies we have Eq.~(\ref{s32}). Considering Eq.~(\ref{rr2}) in the limit $r \rightarrow \infty$, we obtain
\b
\varphi_e^{II} = \frac{\zeta_e}{r} + \tau_e.
\label{r23}
\e
Using Eqs.~(\ref{s32}), (\ref{coe1}) and (\ref{r23}), we find
\b
\begin{array}{ll}
a= \Aetra / (\rho \omega v),\\
b=iM^2\omega \rho \Aetra.
\end{array}
\label{cr2}
\e
We substitute Eq.~(\ref{cr2}) in Eq.~(\ref{aba}) to obtain
\b
\begin{array}{ll}
\Aeinc = -\Aetra\frac{\left(1+M^2\omega^2 \rho^2 v\right)}{2i\rho\omega v},\\
\Aeref =\Aetra\frac{\left(1-M^2\omega^2 \rho^2 v\right)}{2i\rho \omega v},
\end{array}
\label{co3}
\e
which gives us for the absorption cross section, for $l=0$,
\b
\siglf^e =  \frac{\pi}{\omega^2 v^2}\left(\frac{4M^2\omega^2 \rho^2 v}{\left(1+M^2\omega^2 \rho^2 v \right)^2}\right),
\label{lfce}
\e
in the low-frequency regime. Taking $\rho\approx 1$ and considering only the first term in the denominator of Eq. (\ref{lfce}), we are again
left with
\b
\siglf^e = \frac{4\pi M^2}{v} =\frac{\mathcal{A}_{e}}{v} .
\label{lfce2}
\e

\section{Numerical computations\label{sec:numerical}}

In this section we present numerical results for the absorption cross section, obtained by solving the radial equation, Eq.~(\ref{fdr}), numerically. We integrate the radial equation from (close to) the event horizon to a large $r$. By matching the numerical solutions onto the asymptotic forms in Eq.~(\ref{sol}), we obtain the reflection and transmission coefficients and, via Eqs. (\ref{acs}--\ref{pcs}), the absorption cross section of the massive scalar field for the Reissner-Nordstr\"om spacetime.

\begin{figure}[!htb]
\includegraphics[width=\columnwidth]{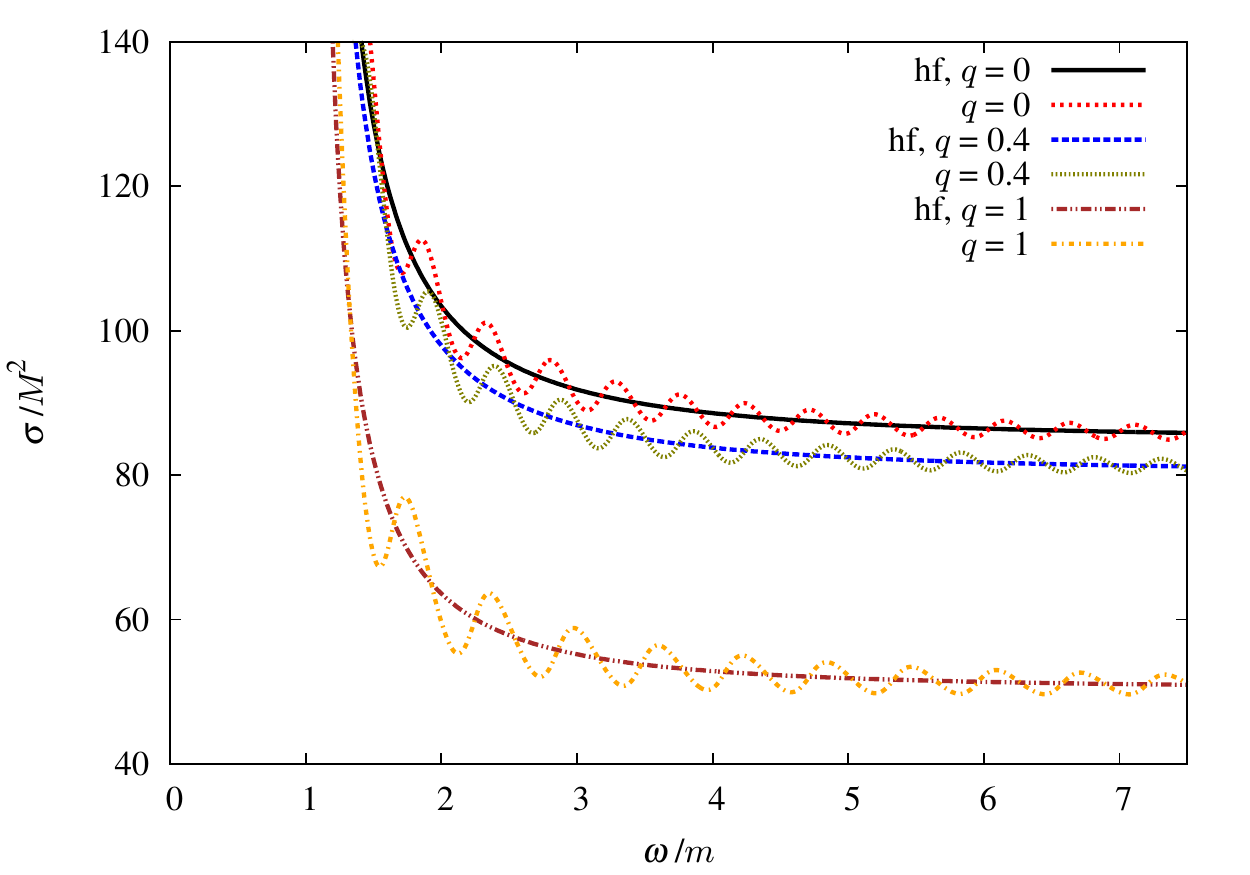}
\caption{Total absorption cross section for $\Mm=0.4$ and for different values of the black hole charge $q$. We also plot the classical (high-frequency) limit $\sighf$ in each case.}
\label{fig:tacsm}
\end{figure}

\begin{figure}[!htb]
\includegraphics[width=\columnwidth]{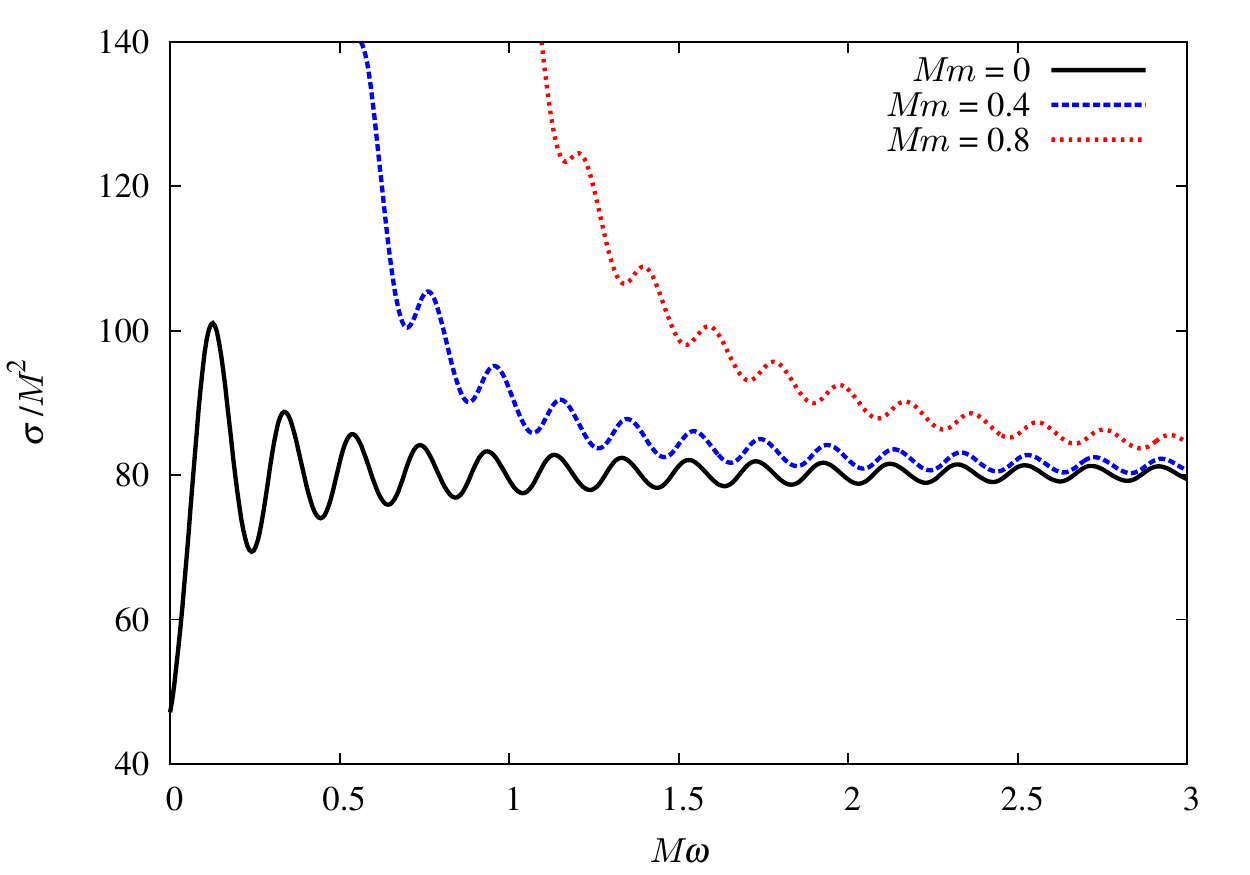}
\caption{Total absorption cross section for $q=0.4$ and different values of $\Mm$.}
\label{fig:tacsq}
\end{figure}

In Fig.~\ref{fig:tacsm} we compare the scalar-wave absorption cross section with the geodesic capture cross section $\sighf = \pi b_c^2$ [cf.~Eq.~(\ref{Fdef})]. We see that $\sigma$ exhibits regular oscillations around $\sighf$. We note that the critical impact parameter $b_c$, and hence also the absorption cross section, diminishes as the the charge-to-mass ratio $q$ increases.  This is in agreement with results for the massless case \cite{Crispino:2009ki}.

In Fig.~\ref{fig:tacsq} we examine the effect of varying the mass of the scalar field $m$. In the case $Mm >Mm_c(q,l)$, the cross section diverges as $1/v^2$ in the limit $\omega \rightarrow m$, as expected from Eq. (\ref{pcs}). As described in Sec.~\ref{sec:lowfreq}, in the very low-frequency regime the cross section instead diverges as $1 / v$. In the high-frequency limit ($\omega \gg 1$ and $\omega / \mass \gg 1$ [or equivalently $v \rightarrow 1$]), Fig.~\ref{fig:tacsq} shows that the massive results converge with their massless counterparts.

\begin{figure}[!htb]
\includegraphics[width=\columnwidth]{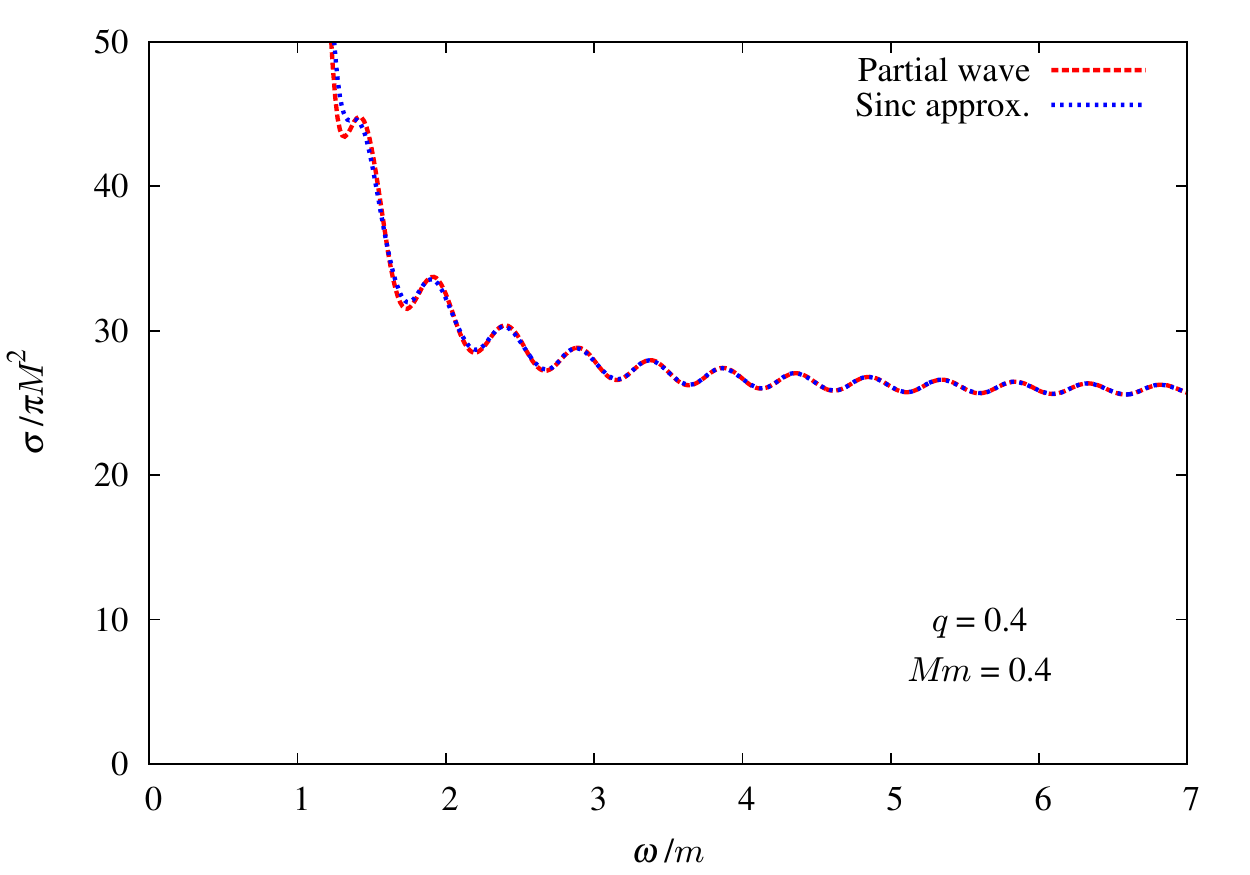}
\caption{Comparison of the `sinc approximation', Eq.~(\ref{eq:sinc}), with numerical results from the partial-wave method, for the case $\Mm = 0.4$, $q = 0.4$. Similar levels of agreement are found for all $q$, in the moderate-to-large $\omega$ regime.}
\label{fig:sinc}
\end{figure}

Figure \ref{fig:sinc} shows that the oscillations in the cross section are well modeled by the sinc approximation in Eq.~(\ref{eq:sinc}). As anticipated [cf. Eq. (\ref{eq:sinc})], the width of the oscillation (in $\omega$) approaches $1/(v b_c)$ in the limit $\omega \rightarrow \infty$.  

\begin{figure}[!htb]
\includegraphics[width=\columnwidth]{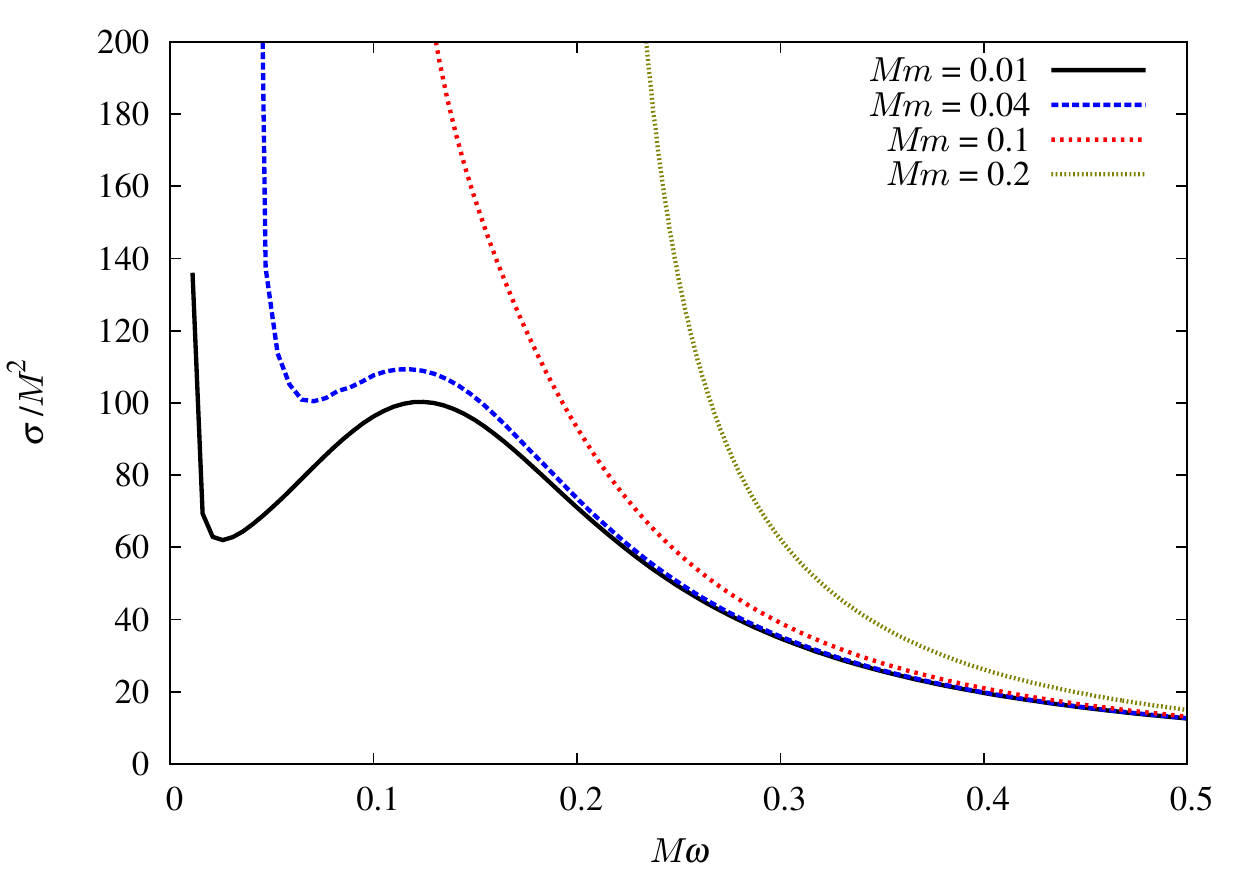}
\caption{Partial absorption cross section for the monopole ($l=0$) mode for charge-to-mass ratio $q=0.4$, and four choices of $\Mm$. Note that the local minimum and maximum disappear as $\Mm$ increases towards $\Mm_c = 0.195$.}
\label{fig:pacsl0}
\end{figure}

In Fig.~\ref{fig:pacsl0} we show the partial absorption cross section for $l=0$ and different values of the mass coupling $\Mm$, above and below the critical mass $\Mm_c$ [for $q=0.4$ and $l=0$, $Mm_c=0.195$]. For $\Mm \ll \Mm_c$ the absorption cross section presents a local minimum and a local maximum. For $\Mm \gtrsim Mm_c$, $\sigma_{l=0}$ becomes a monotonic function of the frequency. 

\begin{figure}[!htb]
\includegraphics[width=\columnwidth]{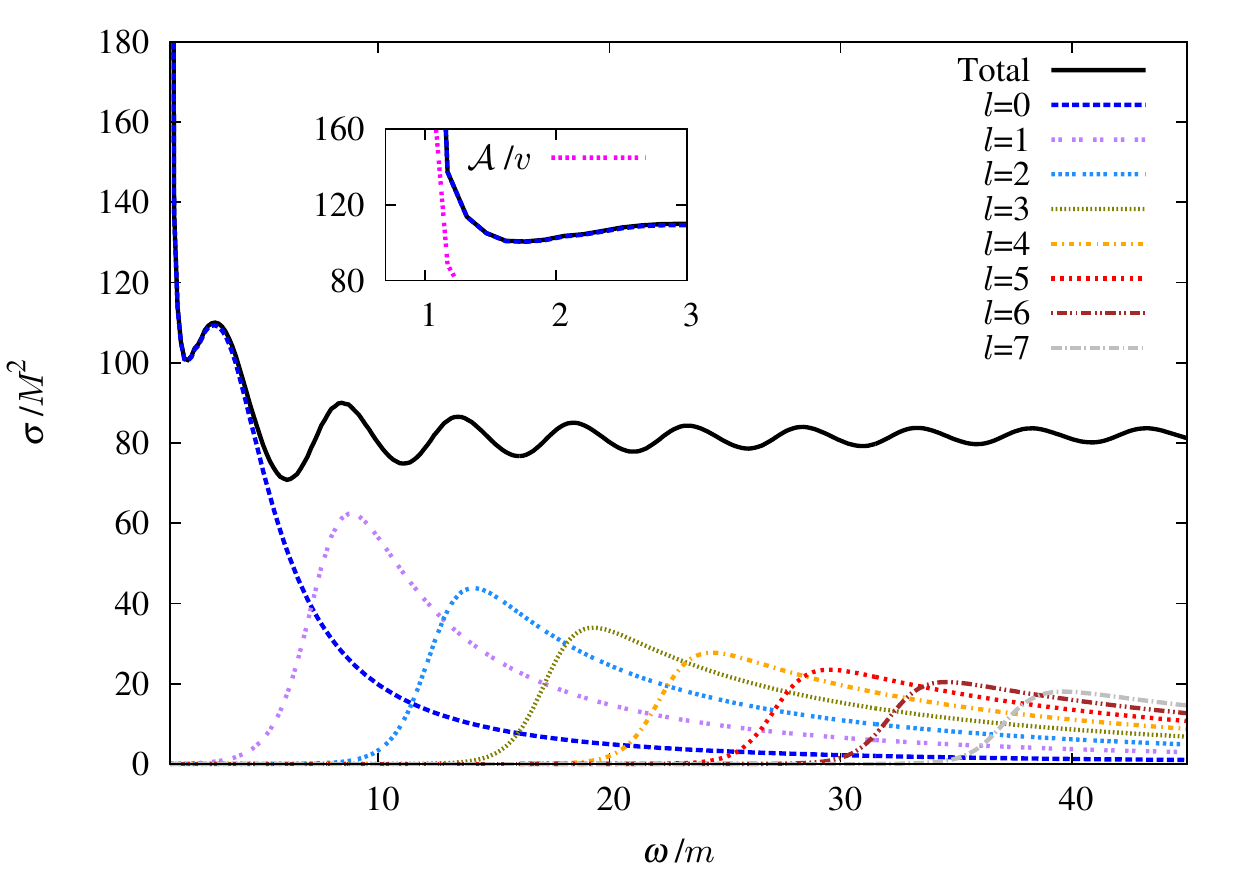}
\caption{Total and partial absorption cross sections for $q=0.4$ and $\Mm=0.04$. The smaller plot shows the low-frequency limit of the absorption cross section.}
\label{fig:pacs:Mm004}
\end{figure}

\begin{figure}[!htb]
\includegraphics[width=\columnwidth]{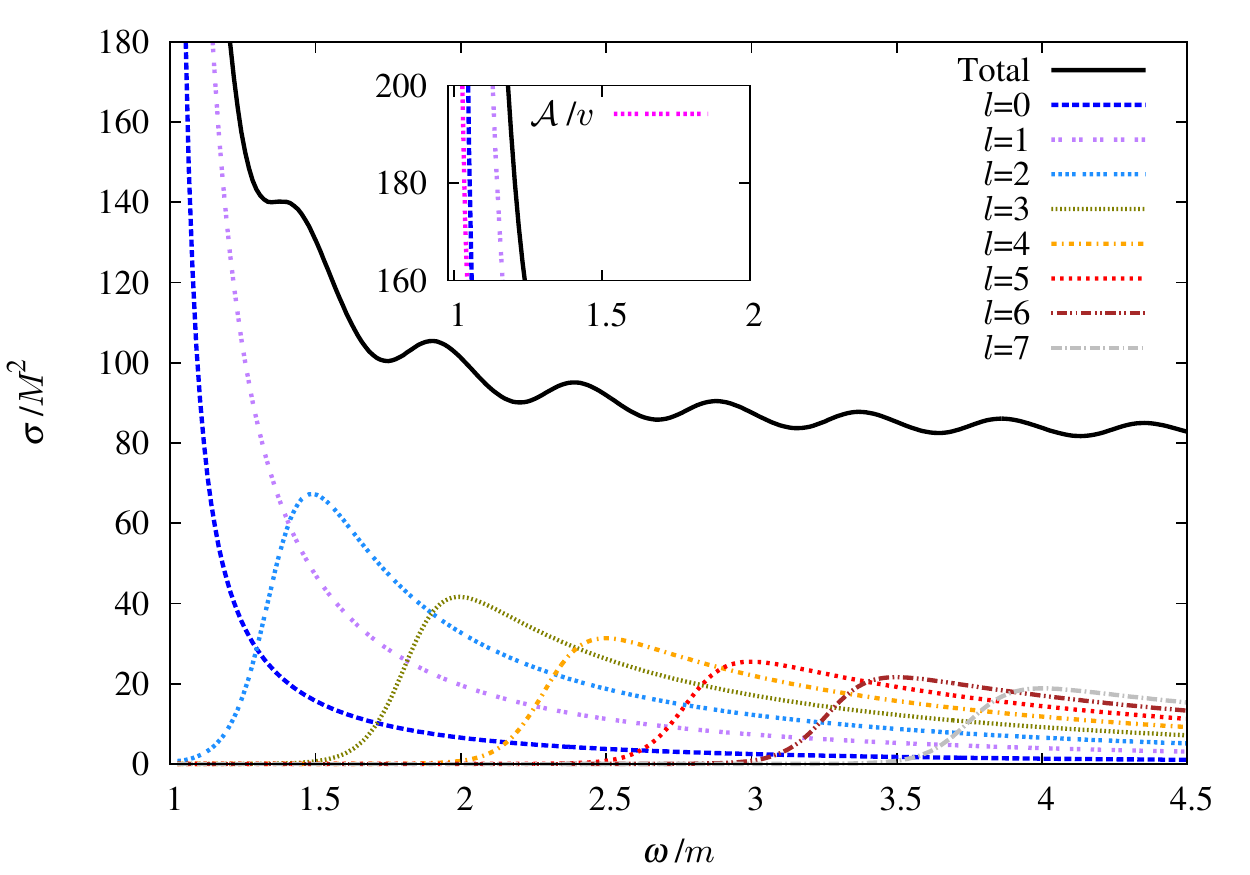}
\caption{Total and partial absorption cross sections for $q=0.4$ and $\Mm=0.4$. The smaller plot shows the low-frequency limit of the absorption cross section.}
\label{fig:pacs:Mm04}
\end{figure}

Figures \ref{fig:pacs:Mm004} and \ref{fig:pacs:Mm04} show the total and partial absorption cross section for $\Mm=0.04$ and $\Mm=0.4$, respectively. For $\Mm=0.04$ we see that the monopole $(l=0)$ gives the main contribution for $\omega/\mass \lesssim 5$. As shown in Sec.~\ref{sec:lowfreq}, in the low-frequency limit the absorption cross section tends to $\mathcal{A}/v$, which diverges as $\omega \rightarrow m$. For $\Mm=0.4$ we see that both the partial cross sections for $l=0$ and $l=1$ diverge in this limit. This occurs because the value of $M m$ in this case is very close to the critical value $Mm_c$ [N.B.~for $q=0.4$ and $l=1$, $Mm_c=0.405$]. We note that, in this case, since $\Mm=0.4$, the low-frequency approximation $A/v$, although still valid for the partial cross section $\sigma_{l=0}$, is not a good approximation for the total low-frequency absorption cross section, as the condition $\Mm \ll 1$ is not fully satisfied and $\sigma_{l=1}$ also diverges as $v \rightarrow 0$.

\begin{figure*}
\centering
\includegraphics[width=\columnwidth]{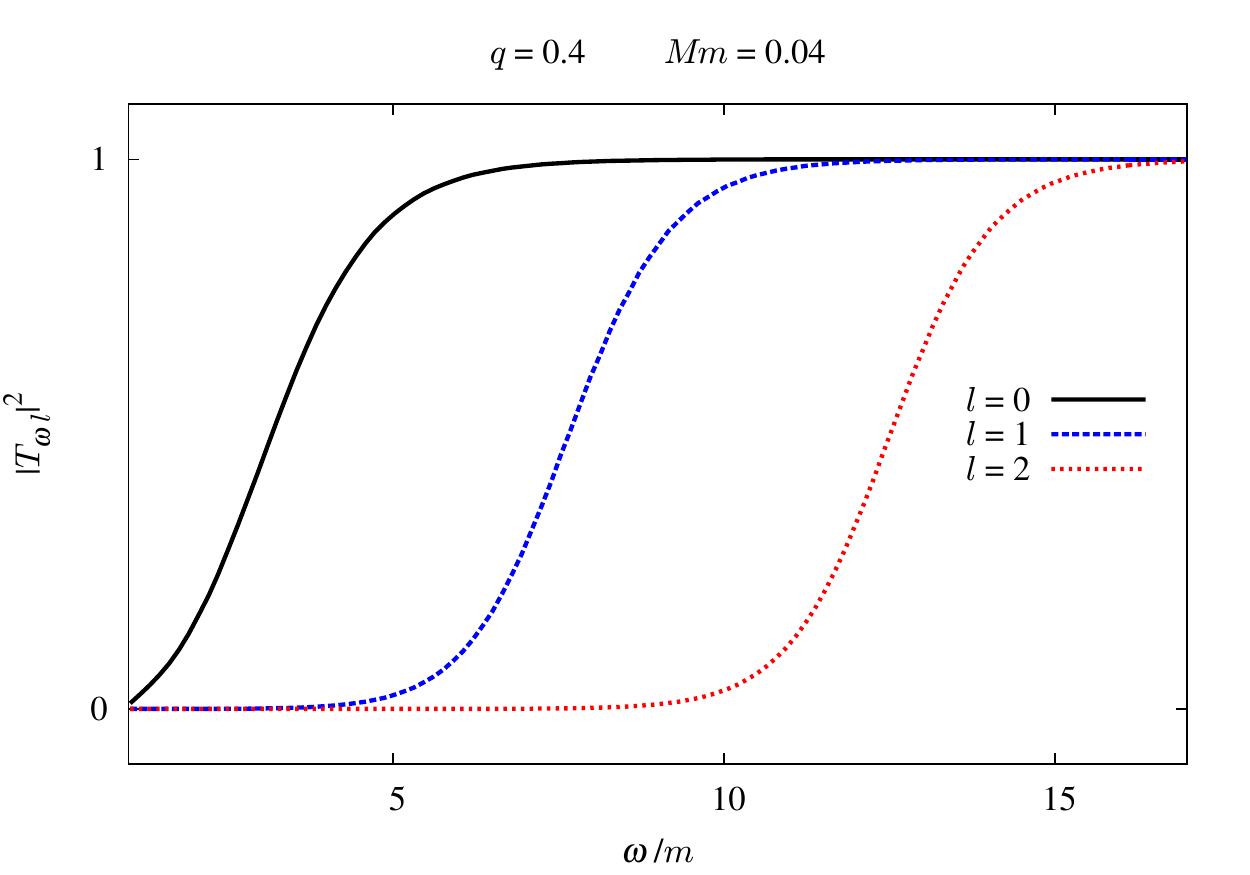}
\includegraphics[width=\columnwidth]{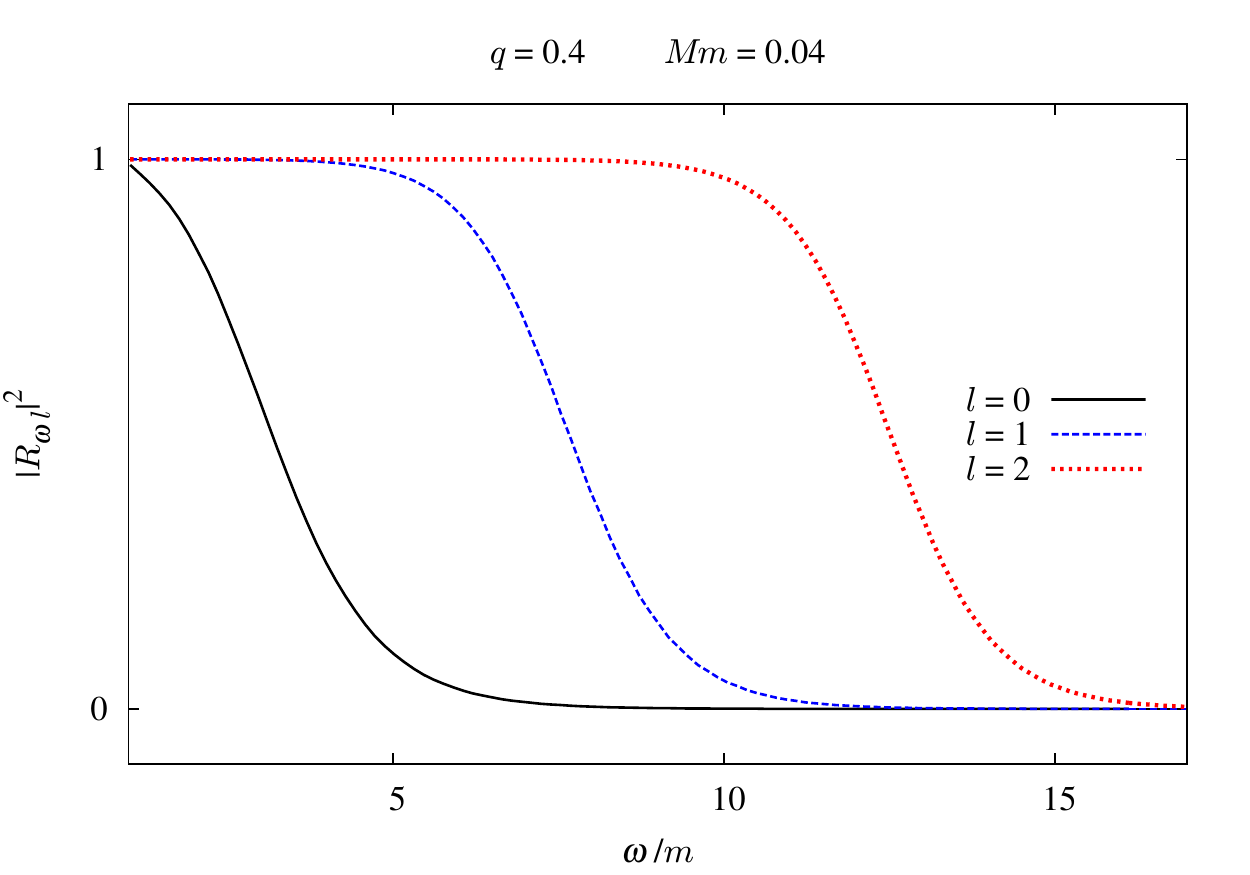}
\includegraphics[width=\columnwidth]{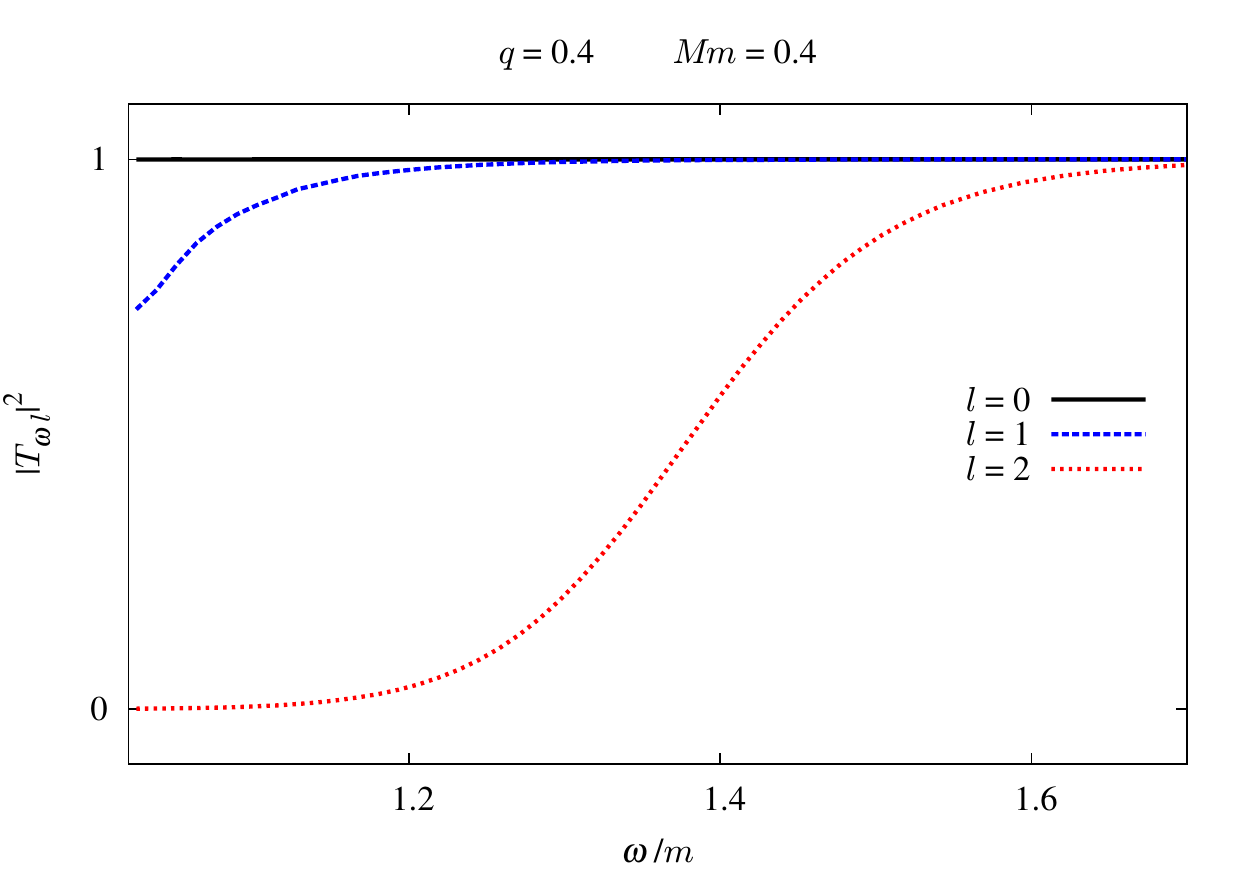}
\includegraphics[width=\columnwidth]{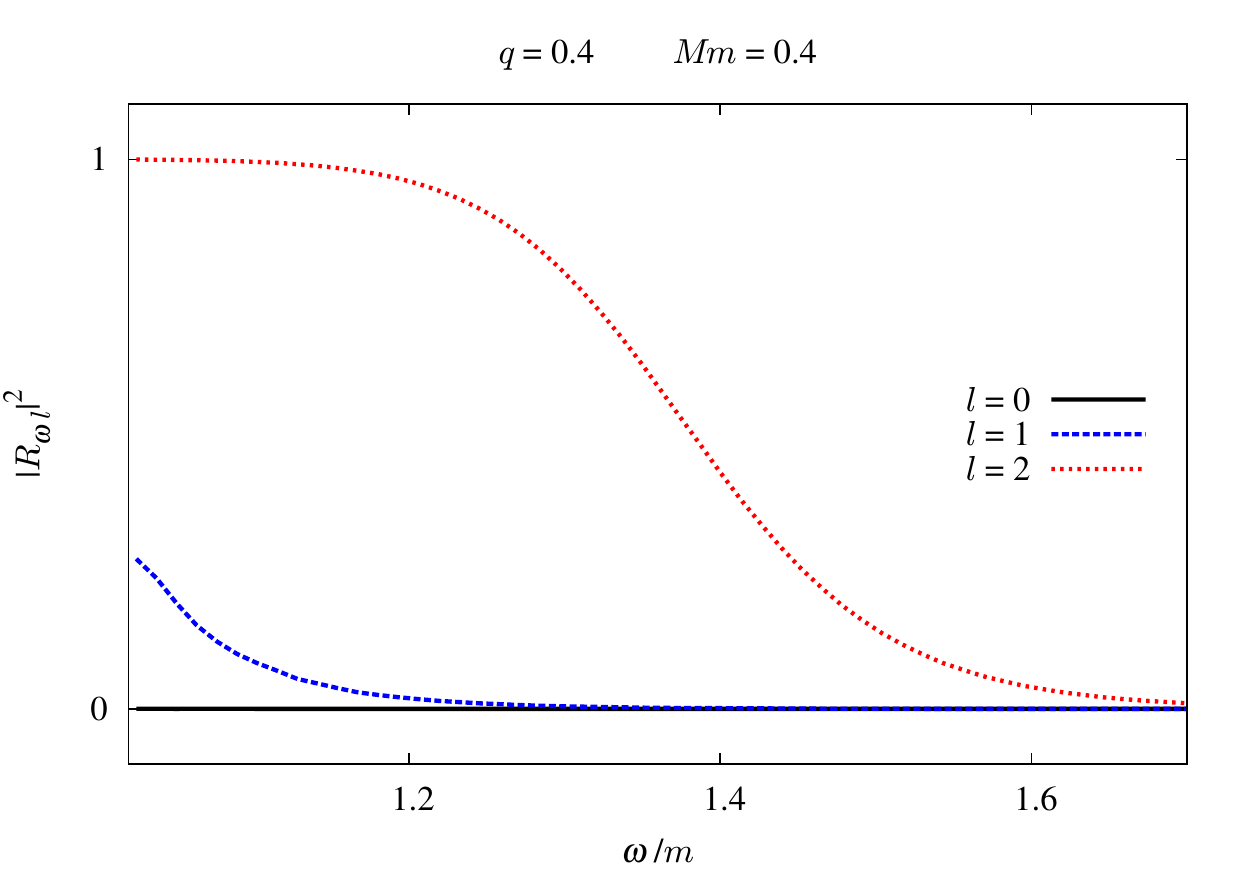}
\caption{Transmission (left plots) and reflection (right plots) coefficients for $q=0.4$ and for mass couplings $\Mm=0.04$ (top plots) and $\Mm=0.4$ (bottom plots), for $l = 0$, $1$ and $2$.}
\label{fig:transrefl}
\end{figure*}

In Fig.~\ref{fig:transrefl} we plot the transmission and reflection coefficients for $\Mm=0.04$ and $\Mm=0.4$ ($q=0.4$). We observe that in the case with smaller mass coupling ($\Mm=0.04$) the transmission  coefficient starts at zero and then goes to unity as $\omega$ increases, for all values of $l$. By contrast, in the case of mass coupling $\Mm=0.4$, the transmission coefficient for $l=0$ is close to the unity for all frequencies. This may be understood by noting that $\Mm > \Mm_c$ in this case, and so there is no effective potential barrier for incident waves; hence near-total absorption is to be expected. 

\begin{figure*}
\includegraphics[width=\columnwidth]{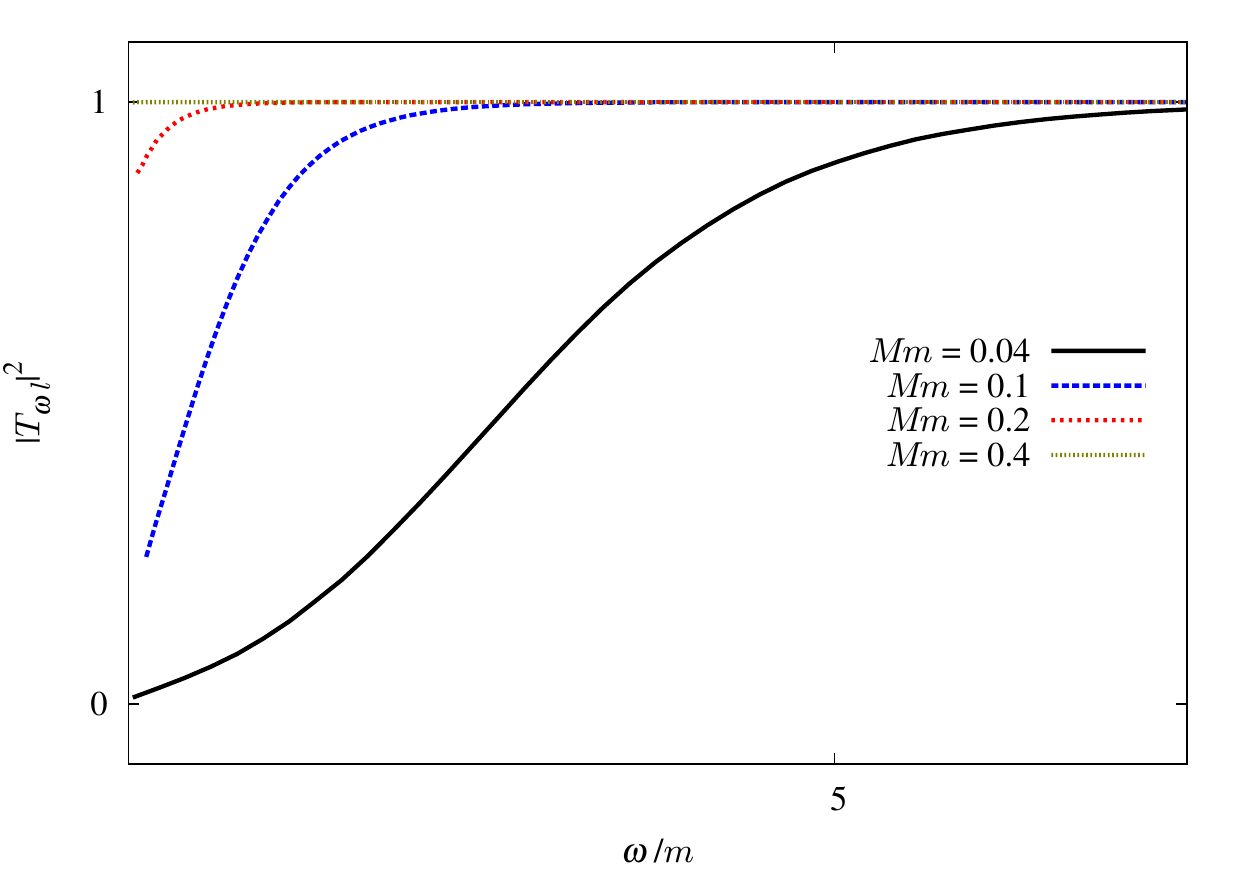}
\includegraphics[width=\columnwidth]{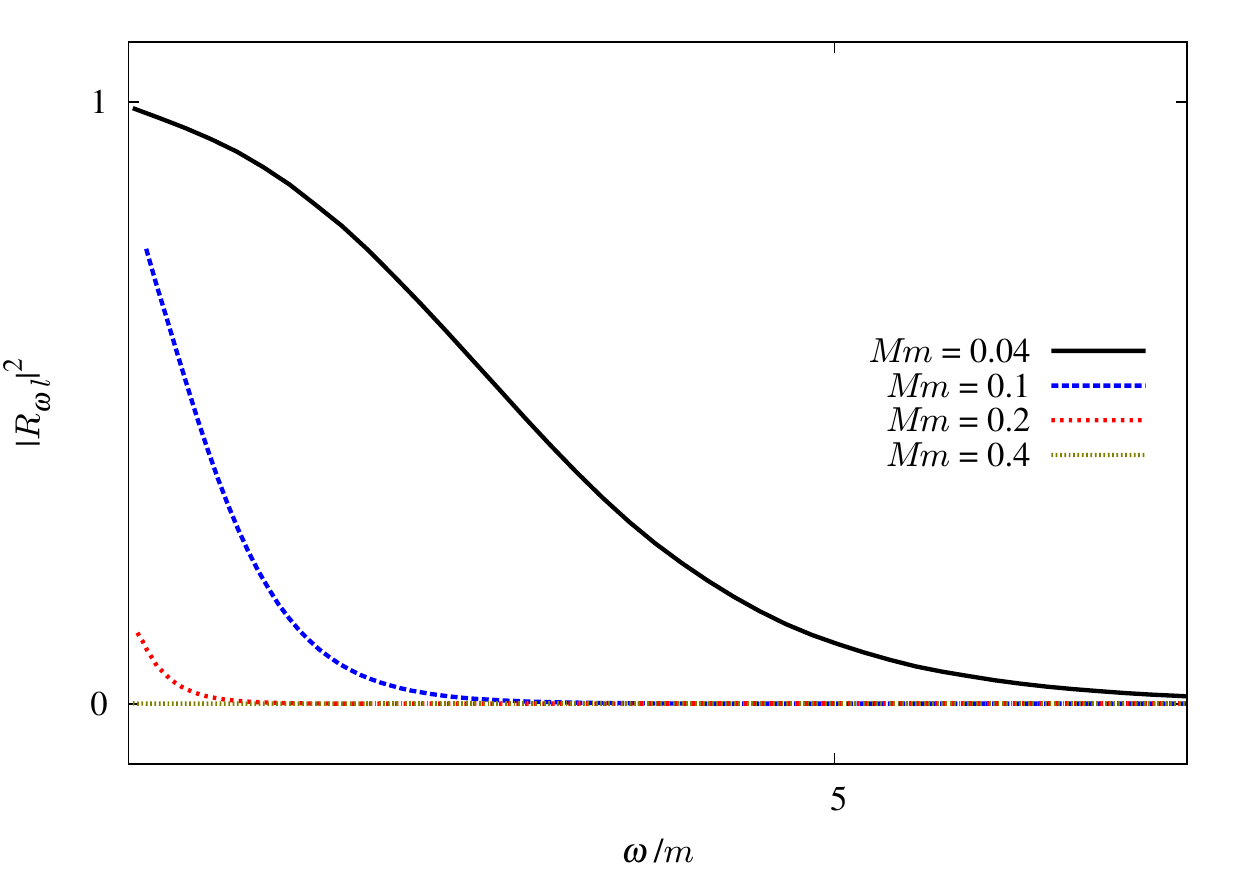}
\caption{Transmission (left) and reflection (right) coefficients for $q=0.4$ and $l=0$, for different choices of $\Mm$.}
\label{fig:transrefl:l0}
\end{figure*}

In Fig.~\ref{fig:transrefl:l0} we show the transmission and reflection coefficients for the monopole ($l = 0$) for a selection of values of the mass coupling. We can see that for $\Mm=0.04$ the transmission coefficient starts near zero, but, as we choose larger values of the mass coupling, the value at $\omega = m$ ($v=0$) increases. Beyond the critical mass $\Mm > \Mm_c = 0.195$ the value at $\omega = m$ is very close to unity.

\section{Final remarks\label{sec:conclusion}}
We have computed the absorption cross section of a massive scalar field by a Reissner-Nordstr\"om black hole for a range of frequencies. We compared our results against (semi-)analytic approximations derived in the high- and low-frequency regimes.

In the moderate-to-high frequency regime, we have verified that the total absorption cross section oscillates around the geodesic capture cross section, as quantitatively described in Sec.~\ref{sec:highfreq}. We have shown that the regular oscillations in the cross section (as a function of frequency) are encapsulated by the `sinc' approximation [Eq.~\eqref{eq:sinc}], which we derived via the complex angular momentum formalism. Following Refs.~\cite{Decanini:2011xi, Decanini:2011xw}, we showed that the properties of the oscillations in the cross section (i.e.~their frequency and amplitude) are set by the frequency and Lyapunov exponent of the unstable orbit in the spacetime at $r=r_c$. 

For small frequencies ($\omega \gtrsim m$), we uncovered distinct possibilities. In the low-frequency limit ($\Mm \ll 1$, $M \omega \ll 1$), absorption is dominated by the monopole, and we established in Sec.~\ref{sec:lowfreq} that $\sigma \sim \siglf = \mathcal{A} / v$, where $\mathcal{A}$ is the area of the event horizon. For $\Mm \ll \Mm_c$, we found that absorption in the monopole exhibits a local minimum and a local maximum (Fig.~\ref{fig:pacsl0}), whereas for $\Mm \gtrsim \Mm_c$ absorption by the monopole increases monotonically as $v \rightarrow 0$. The critical mass $\Mm_c$ increases somewhat with charge-to-mass ratio $q$, as shown in Fig.~\ref{fig:critmass}. For $\Mm > \Mm_c(l)$, the mode $l$ is essentially entirely absorbed by the black hole. Hence, if $\Mm > \Mm_c(l=0)$ then, by Eq.~(\ref{pcs}), the cross section will diverge as $v^{-2}$ (rather than $v^{-1}$) in the limit $v \rightarrow 0$.

A key goal here has been to quantify the effect of both field mass and black hole charge on absorption. We have found that, in general, the effect of the black hole charge is to shift key features of the absorption profile of the Schwarzschild black hole. Let us briefly compare a charged black hole with an uncharged black hole of identical mass. The former appears `smaller' than the latter, in several regards, as the former (i) has a smaller horizon area, (ii) has a smaller critical impact parameter $b_c$, (iii) casts a smaller shadow when illuminated by background radiation, (iv) possesses an unstable circular orbit with a smaller radius (and higher orbital frequency), and (v) exhibits (in general) a smaller scalar-wave absorption cross section, than the latter. These points are interrelated. The critical impact parameter $b_c$ determines the size of the shadow and also the absorption cross section in the high-frequency regime. Via Eq.~(\ref{eq:sinc}), $b_c$ determines the width of the oscillations-with-frequency seen in (e.g.)~Figs.~\ref{fig:tacsm}--\ref{fig:sinc}. The amplitude and decay of these oscillations are set by the critical impact parameter [Eq. (\ref{bc-def})] and the Lyapunov exponent of the unstable circular orbit [Eq. (\ref{eq:lyapunov})], whose dependence on $q$ and $v$ is subtle (see Fig.~\ref{fig:orbparams}). 

The field mass creates qualitatively new effects, leading to (e.g.)~a divergence in the cross section as $v \rightarrow 0$, and total absorption in low multipoles $l+1/2 \lesssim \gamma \Mm$, where the numerical coefficient $\gamma$ may be inferred from Fig.~\ref{fig:critmass}. For any known massive Standard Model fields on a solar-mass black hole spacetime, $\Mm \gg 1$; in such cases, the horizon scale is many orders of magnitude larger than the Compton wavelength of the massive field. However, this is not necessarily true for primordial black holes, or for (posited) ultralight particles such as the axion. To get (e.g.) $\Mm \sim 10^{-2}$, one may have $M \sim 10^8$\,kg in the case of the Higgs boson; $M \sim 10^{11}$\,kg in the case of the neutral pion; or (e.g.)~$M \sim 2 \times 10^{30}$\,kg for an axion of mass $m \sim 10^{-12}$\,eV. 

Let us conclude by speculating on the possible physical relevance of the absorption scenario. In the foreseeable future, it is possible that observations in the electromagnetic spectrum of black hole `shadows' will become feasible \cite{Li:2013jra}. Such measurements would allow one to probe the absorption cross section in the high frequency regime, $\sighf$ (see Sec.~\ref{sec:highfreq}). By combining $\sighf$ with an independent measurement of the black hole's mass, one may attempt to deduce the black holes charge-to-mass ratio $q$. In practice, it seems likely that astrophysical black holes have negligible charge but significant angular momentum, so the focus will be on deducing the spin parameter instead. We note that rotating black holes generate asymmetric shadows, which may give additional ways to break degeneracies in parameter space \cite{Macedo:2013afa}.  If it becomes possible to sample the absorption cross section across a frequency band, then quantitative studies such as this will play a role in cleanly extracting all key parameters ($M$, $\Mm$ and $q$, say). Notwithstanding this possibility, the results herein represent a further step towards a quantitative understanding of the interaction of black holes with surrounding fields.

\section{Acknowledgments}

We would like to acknowledge Conselho Nacional de Desenvolvimento
Cient\'ifico e Tecnol\'ogico (CNPq), Coordena\c{c}\~ao de Aperfei\c{c}oamento de Pessoal de N\'ivel Superior (CAPES) and Marie Curie action
NRHEP-295189- FP7-PEOPLE-2011-IRSES for partial financial
support. LC acknowledges partial support from Abdus Salam International
Centre for Theoretical Physics through the Associates Scheme. CB is grateful to Caio F. B. Macedo for profitable discussions.

\bibliography{bodc.bib}
\end{document}